\documentclass[paper, 12pt, letterpaper]{JHEP} 
\usepackage{amsmath}
\usepackage{amssymb}

\def\Bbb{\mathbb} \def\C{{\Bbb C}} \def\R{{\Bbb R}} \def\Z{{\Bbb Z}}

 \def\Re {\bb{Re}}  

    \def\tr{\operatorname{tr}}\def\Tr{\operatorname{Tr}}
\def\str{\operatorname{str}}
\def\Str{\operatorname{Str}}

\def\Re{{\rm Re\,}} \def\tr{{\rm tr\, }} 
 \def\Tr{{\rm Tr\, }} 
   \def\deg{{\rm deg}} \def\rk{{\rm rk}} \def\det{{\rm det}}

\def\id{\protect{{1 \kern-.28em {\rm l}}}}

\newcommand{\be}{\begin{equation}} \newcommand{\ee}{\end{equation}}
\newcommand{\bea}{\begin{eqnarray}} \newcommand{\eea}{\end{eqnarray}}
\newcommand{\beann}{\begin{eqnarray*}}
  \newcommand{\eeann}{\end{eqnarray*}}
\newcommand{\bfig}{\begin{figure}} \newcommand{\efig}{\end{figure}}
\newcommand{\nn}{\nonumber}
\newcommand{\ba}{\begin{array}}\newcommand{\ea}{\end{array}}

\newtheorem{Proposition}{Proposition}[section]

\newtheorem{Theorem}{Theorem}[section]
\newtheorem{Lemma}{Lemma}[section]
\newtheorem{Corrolary}{Corrolary}[section]

\newcommand{\bp}{\begin{Proposition}}
  \newcommand{\ep}{\end{Proposition}}
\newcommand{\bt}{\begin{Theorem}} \newcommand{\et}{\end{Theorem}}
\newcommand{\bl}{\begin{Lemma}} \newcommand{\el}{\end{Lemma}}
\newcommand{\bc}{\begin{Corrolary}} \newcommand{\ec}{\end{Corrolary}}

   \def\ep{\eps}

\author{
Manfred Herbst${}^a$, Calin-Iuliu Lazaroiu${}^b$ \\}
\author{~\\
     ${}^a$Department of Physics, CERN\\
     Theory Division\\
     CH-1211 Geneva 23\\
     Switzerland\\
     Manfred.Herbst@cern.ch\\{~}\\
     ${}^b$5243 Chamberlin Hall\\
     University of Wisconsin at Madison\\
     1150 University Ave\\
     Madison, Wisconsin 53706, USA\\
     calin@physics.wisc.edu\\
    }

\title{Localization and traces in open--closed topological Landau--Ginzburg models}

\abstract{We reconsider the issue of localization in open-closed B-twisted
Landau-Ginzburg models with arbitrary Calabi-Yau target. Through careful
analsysis of zero-mode reduction, we show that the closed model allows for a
one-parameter family of localization pictures, which generalize the standard
residue representation. The parameter $\lambda$ which indexes these pictures
measures the area of worldsheets with $S^2$ topology, with the residue
representation obtained in the limit of small area. In the boundary sector, we
find a double family of such pictures, depending on parameters $\lambda$ and
$\mu$ which measure the area and boundary length of worldsheets with disk
topology. We show that setting $\mu=0$ and varying $\lambda$ interpolates
between the localization picture of the B-model with a noncompact target space
and a certain residue representation proposed recently. This gives a complete
derivation of the boundary residue formula, starting from the explicit construction of the
boundary coupling. We also show that the various localization pictures are
related by a semigroup of homotopy equivalences.}

\preprint{MAD-TH-04-3\\  
          CERN--PH--TH/2004-070}

\begin{document}

\tableofcontents

\pagebreak

\vskip .6in

\section{Introduction}
\label{intro}

Closed topological Landau-Ginzburg models \cite{Vafa_LG} have been a useful
testing ground for string theory. They make direct contact with
the topological sector of rational conformal field theories 
through the Landau-Ginzburg approach to minimal models, and arise as phases of
$N=2$ string compactifications \cite{Witten_phases}.
Moreover, they give explicit realizations of the 
WDVV equations and examples of Frobenius manifolds, and have interesting relations with 
singularity theory.

In a similar vein, one can expect to learn important lessons about open
string theory by studying topological Landau-Ginzburg models in the presence
of D-branes (see \cite{us, fractional} for some recent results in this direction). 
While basic D-brane constructions were considered by many authors
(see \cite{Hori} and references therein), a systematic study has been
hampered by the lack of a reasonably general description of the boundary
coupling (the "Warner problem" \cite{Warner}).

Progress in removing this obstacle was made recently in  \cite{Kontsevich, Kap1,  Lerche,
  Kap2, Kap3} (see also \cite{Orlov}). These papers proposed a solution of the Warner problem for
  B-twisted Landau-Ginzburg models with target space $\C^n$ and for particular 
families
  of D-branes described by superbundles whose rank is constrained to be a
  power of two. In a slightly modified form, this 
  solution was generalized in \cite{coupling} by removing unnecessary 
  assumptions, thus giving the general form of the
  relevant boundary coupling.

As in the closed string case, it is natural to translate the
physical data of open Landau-Ginzburg models into the language of 
singularity theory. For the closed string sector, a crucial step in this regard is
the localization formula of \cite{Vafa_LG}, which relates topological field
theory correlators to residues (see, for
example, \cite{Griffiths}). An open string version of this formula was  proposed in \cite{Kap2}, 
though a complete derivation based on the microscopic boundary coupling was not given. 

Given the boundary coupling constructed in \cite{coupling}, we shall re-consider this issue in the 
more general set-up of \cite{Labastida}, and give a complete derivation of
this localization formula. 
Another purpose of this paper is to extend the open and closed localization
formulae in a manner which reflects the basic intuition 
\cite{coupling} that the B-branes of Landau-Ginzburg models are the
result of tachyon condensation between the elementary branes of the B-type
sigma model, with tachyon condensation driven by the Landau-Ginzburg superpotential. 
As we shall show somewhere else, this allows one to make contact with the
string field theory approach advocated in a different context in 
\cite{Witten_CS, CIL2, CIL3, CIL4, CIL5, CIL6, CIL7, CIL8, CIL9, CIL10, Diaconescu}.

Perhaps surprisingly, this is non-trivial to
achieve already for the closed string sector. Indeed, the usual on-shell descriptions
of the space of bulk  observables differ markedly
between the two models. In the B-twisted sigma model, this space is described
as the ${\bar \partial}$-cohomology of the algebra of $(0,p)$-forms on the
target space, valued in holomorphic polyvector fields. In the Landau-Ginzburg
case, the space of on-shell bulk observables can be identified with the Jacobi ring
of the Landau-Ginzburg superpotential. As we shall see below, these two
descriptions are related in a subtle manner, namely by a one-parameter
family of "localization pictures" which interpolates between them. The
existence of such a family will be established by refining 
the localization argument of \cite{Vafa_LG}. 
The different pictures are indexed by a parameter $\lambda$, 
which roughly measures the area of 
worldsheets with $S^2$ topology.
The B-model description of the algebra of observables
arises in the limit when the worldsheet is collapsed to a point, while the 
Jacobi realization is recovered for very large areas. More
precisely, one can construct an off-shell model for each localization 
picture, with a reduced BRST operator whose cohomology reproduces the space
of on-shell bulk observables. The various pictures 
are related by a "homotopy flow", i.e. a one-parameter semigroup of operators 
homotopic to the identity. 
This flow induces the trivial action on BRST cohomology, thus identifying 
on-shell data between different pictures. 

Extending this construction to the boundary sector, we find a similar
description. Namely, we shall 
construct a family of localization pictures indexed by {\em two}
parameters $\lambda$ and $\mu$, which -- when real -- measure the area of a worldsheet with disk
topology and the length of its boundary. 
Taking $\mu=0$, the standard realization of observables in the open B-model
arises  for $\lambda\rightarrow 0$, while the LG description and residue formula 
of \cite{Kap2} are obtained in the opposite limit $\lambda\rightarrow +\infty$. The parameter
$\mu$ plays an auxiliary role, related to a certain boundary term which was
not included in \cite{coupling} since it is not essential for the
topological model. Contrary to previous proposals, we show that this
parameter can be safely set to zero, without affecting the localization data. 
Physically, localization on the critical set of $W$ is controlled by the 
bulk parameter $\lambda$, and is completely independent of the choice of
$\mu$. In fact, one must set $\mu=0$ in order to recover
\footnote{The authors of \cite{Kap2} propose a different limit, namely
$\mu\rightarrow +\infty$ with $\lambda=0$. It does not seem possible to achieve
localization on the critical set of $W$ in this limit.} the residue representation of 
\cite{Kap2}.

The paper is organized as follows. In Section 2, we review the bulk Lagrangian
and some of its basic properties, following \cite{Labastida}. In Section 3, we
discuss localization in the bulk sector. Through careful 
analysis, we show that one can localize on the
zero modes of the {\em sigma model} action, namely that part of the bulk action
which is independent of the Landau-Ginzburg superpotential $W$. This gives the
one-parameter family of localization pictures. We also give 
the geometric realization of these pictures, and the homotopy flow connecting them. 
In Section 5, we recall the boundary coupling given in  \cite{coupling} and
adapt it by adding a supplementary term
also suggested in \cite{Kap1, Lerche, Kap2} for a special case. This assures that
the coupling preserves a full copy of the $N=2$ topological algebra, when the
model is considered on a flat strip. Since the second generator of this
algebra is gauged when considering the model on a curved Riemann surface, this
condition does not play a fundamental role for unintegrated amplitudes, 
but the modified coupling is useful for comparison with \cite{Kap2}.
Section 5 constructs the boundary observables and correlators, and explains
how our model for the boundary  BRST operator arises in this approach. In Section 6, we
discuss localization in the boundary sector. As for the bulk, we proceed by
reducing to zero modes of the sigma model action. This gives a family 
of representations depending on two parameters $\lambda$ and $\mu$.
The second of these weights the contribution arising from the supplementary
boundary term. After describing boundary homotopy flows and
the associated geometric realization, we construct an off-shell representative
for the bulk-boundary map of \cite{CIL1} (see also \cite{Moore_Segal, Moore}), 
and use it to recover (an extension
of) the localization formula proposed in \cite{Kap2}, by taking the limit
$\lambda\rightarrow +\infty$ with $\mu=0$\footnote{In this limit, the
  supplementary term introduced in Section 5 does not contribute, so 
one can use the simplified boundary coupling of \cite{coupling}.}. 
Section 7 presents our conclusions. The Appendix gives the boundary 
conditions for the general D-brane coupling.

\section{The bulk action}
\label{bulk}

The general formulation of closed B-type topological Landau-Ginzburg models
was given in \cite{Labastida}, extending the work of
\cite{Vafa_LG}. We take the target space to be a Calabi-Yau manifold $X$,
with the Landau-Ginzburg potential a holomorphic function $W\in H^0({\cal
O}_X)$.  Since any holomorphic function on a compact complex manifold is
constant, we shall assume that $X$ is non-compact.  
In the on-shell formulation, 
the Grassmann even worldsheet fields are 
the components $\phi^i$, $\phi^{\bar i}$ of the map $\phi:\Sigma \rightarrow
X$, while the G-odd fields are sections $\eta$, $\theta$ and $\rho$ of the
bundles $\phi^*({\bar T}X)$, $\phi^*(T^*X)$ and $\phi^*(TX)\otimes {\cal
T}^*\Sigma$ over the worldsheet $\Sigma$. Here ${\cal T}^*\Sigma$ is the
complexified cotangent bundle to $\Sigma$, while $TX$ and ${\bar T}X$ are the
holomorphic and antiholomorphic components of the complexified tangent bundle
${\cal T}X$ to $X$. This agrees with the on-shell field content of 
the B-twisted sigma model \cite{Witten_mirror}.

\subsection{Action and BRST transformations}

To write the bulk action, we introduce new fields $\chi, {\bar
  \chi} \in \Gamma(\Sigma, \phi^*({\bar T}X))$ by the relations:
\bea
\eta^{\bar i}&=&\chi^{\bar i}+{\bar \chi}^{\bar i}\\
\theta_i&=&G_{i{\bar j}}(\chi^{\bar j}-{\bar \chi}^{\bar j})~~.
\eea
We shall also use the quantity $\theta^{\bar i}=G^{{\bar i}j}\theta_j$.

As in \cite{Labastida}, it is convenient to use an off-shell
realization of the BRST symmetry. For
this, consider an auxiliary G-even field ${\tilde F}$ transforming as a section of
$\phi^*({\cal T}X)$. Then the BRST transformations are:
\bea
\label{BRST}
\delta \phi^i=0~~&,&~~\delta \phi^{\bar i}=\chi^{\bar i}+{\bar \chi}^{\bar
  i}=\eta^{\bar i}\nn\\
\delta \chi^{\bar i}={\tilde F}^{\bar i}-\Gamma^{\bar
  i}_{{\bar j}{\bar k}}{\bar \chi}^{\bar j}\chi^{\bar k}~~&,&~~\delta {\bar \chi}^{\bar
  i}=-{\tilde F}^{\bar i}+\Gamma^{\bar i}_{{\bar j}{\bar
  k}}{\bar \chi}^{\bar j}\chi^{\bar k}~~\nn\\
\delta \rho^i_\alpha&=&2\partial_\alpha \phi^i~~\\
\delta {\tilde F}^i=i\varepsilon^{\alpha\beta} \left[D_\alpha
  \rho^i_\beta+\frac{1}{4}R^i_{j{\bar l}k}(\chi^{\bar l}+{\bar \chi}^{\bar l})
\rho^j_\alpha\rho^k_\beta\right]~~&,&~~\delta
{\tilde F}^{\bar i}=\Gamma^{\bar i}_{{\bar j}{\bar k}}{\tilde F}^{\bar j}(\chi^{\bar k}+{\bar \chi}^{\bar k})~~.\nn
\eea
These transformations are independent of
  $W$. Moreover, the transformations of $\phi$, $\eta$ and $\rho$ 
do not involve the auxiliary fields. In particular, we have  $\delta \eta^{\bar i}=0$.
These observations will be used in Section 4. 

Let us pick a Riemannian metric $g$ on the  worldsheet. The bulk action
of \cite{Labastida} takes the form:
\be
\label{Sbulk_decomp}
S_{bulk}=S_B+S_W
\ee
where:
\bea
\label{S_B}
\!\!\!\!\!S_B&=&\int_\Sigma {d^2\sigma \sqrt{g}}{\Big[ G_{i{\bar
j}}\left(g^{\alpha\beta} \partial_\alpha \phi^i \partial_\beta \phi^{\bar j}
-i\varepsilon^{\alpha\beta}\partial_\alpha \phi^i\partial_\beta\phi^{\bar
j}-\frac{1}{2}g^{\alpha\beta}\rho_\alpha^i D_\beta \eta^{\bar
j}-\frac{i}{2}\varepsilon^{\alpha\beta} \rho_\alpha^i D_\beta \theta^{\bar
j}-{\tilde F}^i{\tilde F}^{\bar j}\right)}\nn\\
\!\!\!\!&+&\frac{i}{4}\varepsilon^{\alpha\beta}R_{i{\bar l}k{\bar
j}}\rho^i_\alpha{\bar \chi}^{\bar l}\rho^k_\beta\chi^{\bar j}\Big] 
\eea
is the action of the B-twisted sigma model and $S_W=S_0+S_1$ is the potential-dependent term, with:
\bea
S_0&=&-\frac{i}{2} \int_{\Sigma}{d^2\sigma \sqrt{g} \left[D_{\bar i}
\partial_{\bar j}{\bar W} \chi^{\bar i}{\bar \chi}^{\bar j}-(\partial_{\bar
i}{\bar W}){\tilde F}^{\bar i}\right]}\\
S_1&=&-\frac{i}{2}\int_{\Sigma}{d^2\sigma \sqrt{g} \left[(\partial_i W){\tilde
F}^i + \frac{i}{4} \varepsilon^{\alpha\beta}D_i\partial_jW
\rho_\alpha^i\rho_\beta^j\right]}~~.
\eea
The quantity $\varepsilon^{\alpha\beta}=\frac{\epsilon^{\alpha\beta}}{\sqrt{g}}$ is
the Levi-Civita tensor, while $\epsilon^{\alpha\beta}$ is the associated density. We
have rescaled the Landau-Ginzburg potential $W$ by a factor of $\frac{i}{2}$
with respect to the conventions of \cite{Labastida} (the conventions for the
target space Riemann tensor and covariantized worldsheet derivative $D_\alpha$
are unchanged).  In $S_W$, we separated the term depending on $W$ from that
depending on its complex conjugate.

As shown in \cite{Labastida}, the topological sigma model action 
(\ref{S_B}) is BRST exact on closed Riemann surfaces.  Since in this
paper we shall allow $\Sigma$ to have a nonempty boundary, we must be careful
with total derivative terms. Extending the computation of \cite{Labastida} to
this case, we find:
\be
\label{exactS_B}
S_B+s=\delta V_B
\ee
where:
\be
V_B:=\int_{\Sigma}{d^2\sigma \sqrt{g} G_{i{\bar
  j}}\left(\frac{1}{2}g^{\alpha\beta}\rho^i_\alpha\partial_\beta \phi^{\bar
  j}-\frac{i}{2}\varepsilon^{\alpha\beta}\rho^i_\alpha
  \partial_\beta\phi^{\bar j}-{\tilde F}^i\chi^{\bar j}\right)}
\ee
and:
\be
\label{s}
s:=i\int_{\Sigma}{d^2\sigma
      \sqrt{g}\varepsilon^{\alpha\beta}\partial_\alpha(G_{{\bar i}j}\chi^{\bar
      i}\rho^j_\beta)}=i\int_\Sigma{d(G_{{\bar i}j}\chi^{\bar i}\rho^j)}~~.
\ee
Since total derivative  terms do not change physics on
closed Riemann surfaces, we are free to redefine the bulk 
sigma-model action by adding (\ref{s}) to $S_B$:
\be
{\tilde S}_B:=S_B+s=\delta V_B~~.
\ee
Accordingly, we shall use the modified bulk Landau-Ginzburg action:
\be
\label{tilde_S_bulk}
{\tilde S}_{bulk}=S_{bulk}+s={\tilde S}_B+S_0+S_1~~.
\ee
It is not hard to check that the term $S_0$ is BRST exact:
\be
\label{exactS_0}
S_0=\delta V_0
\ee
where:
\be
V_0=\frac{i}{4}\int_{\Sigma}{d^2\sigma \sqrt{g}\theta^{\bar i} \partial_{\bar
i}{\bar W}}~~.
\ee
Equations (\ref{exactS_B}) and (\ref{exactS_0}) are local, i.e. they hold for
the associated Lagrange densities without requiring integration by parts. Thus
both of these relations can be applied to bordered Riemann surfaces.

Since the boundary term (\ref{s}) is independent of the worldsheet metric, 
the bulk stress energy tensor has the form given in \cite{Labastida}:
\bea
T_{\mu \nu}&=&\frac{1}{2\sqrt{g}}\frac{\delta {\tilde S}_{bulk}}{\delta g^{\mu
    \nu}}=\frac{1}{2}G_{i{\bar j}}\left[\partial_\mu \phi^i\partial_\nu \phi^{\bar j}+
\partial_\nu \phi^i\partial_\mu \phi^{\bar j}-\frac{1}{2}\left(\rho_\mu^iD_\nu
\eta^{\bar j}+\rho_\nu^i D_\mu \eta^{\bar j}\right)\right]-\\
&-&\frac{1}{2}g_{\mu \nu}\left[G_{i{\bar j}}g^{\alpha\beta}(\partial_\alpha\phi^i\partial_\beta\phi^{\bar
    j}-\frac{1}{2} \rho_\alpha^i D_\beta \eta^{\bar j}-{\tilde F}^i{\tilde
    F}^{\bar j})-\frac{i}{2}\partial_i W {\tilde
    F}^i+\frac{i}{2}\partial_{\bar i}{\bar W}{\tilde
    F}^{\bar i}-\frac{i}{4}D_{\bar i}\partial_{\bar j}{\bar W}\theta^{\bar
    i}\eta^{\bar j}\right]~~.\nn
\eea
As explained in \cite{Labastida}, $T_{\mu\nu}$ is BRST exact only modulo
the equations of motion for the auxiliary fields:
\bea
\label{F_eom}
{\tilde F}^i=\frac{i}{2}G^{i{\bar j}}\partial_{\bar j}{\bar W}~~,~~
{\tilde F}^{\bar i}=-\frac{i}{2}G^{{\bar i}j}\partial_j W~~.
\eea
Imposing these equations, one finds:
\bea
T^{os}_{\mu\nu}&=&\frac{1}{2}G_{i{\bar j}}\left[\partial_\mu \phi^i\partial_\nu \phi^{\bar j}+
\partial_\nu \phi^i\partial_\mu \phi^{\bar j}-\frac{1}{2}\left(\rho_\mu^iD_\nu
\eta^{\bar j}+\rho_\nu^i D_\mu \eta^{\bar j}\right)\right]-\\
&-&\frac{1}{2}g_{\mu \nu}\left[G_{i{\bar j}}g^{\alpha\beta}(\partial_\alpha\phi^i\partial_\beta\phi^{\bar
    j}-\frac{1}{2} \rho_\alpha^i D_\beta \eta^{\bar j})+\frac{1}{4}G^{i{\bar j}}\partial_i W\partial_{\bar
    j}{\bar W}-\frac{i}{4}D_{\bar i}\partial_{\bar j}{\bar W}\theta^{\bar
    i}\eta^{\bar j}\right]~~.\nn
\eea
This obeys the BRST exactness condition \cite{Labastida}:
\be
\label{T_exact}
T^{os}_{\mu \nu}=\delta G_{\mu \nu}
\ee
where\footnote{The formula given for $G_{\mu \nu}$ in \cite{Labastida} seems
  to be missing a global prefactor of $-\frac{1}{2}$.}:
\be
\label{G_mu_nu}
G_{\mu\nu}=\frac{1}{4}\Big[G_{i{\bar j}}(\rho_\mu^i\partial_\nu \phi^{\bar j}+
\rho_\nu^i\partial_\mu \phi^{\bar j})-g_{\mu
  \nu}(G_{i{\bar j}}g^{\alpha\beta}\rho_\alpha^i\partial_\beta\phi^{\bar
  j}+\frac{i}{2}\theta^{\bar i}\partial_{\bar i}{\bar W})\Big]~~.
\ee
On an infinite flat cylinder, the supercharges: 
\be
\label{G_mu}
G_\mu:=\int{d\sigma_1 G_{0\mu}} 
\ee
generate symmetries $\delta_\mu=\{G_\mu, \cdot\}_P$ which together with $\delta=\{Q,\cdot\}_P$
and a supplementary nilpotent transformation $\delta':=\{M,\cdot\}_P$ 
form the topological algebra of
\cite{Labastida} (here  $\{\cdot, \cdot\}_P$ is the Poisson bracket of the
Hamiltonian formulation). When placing the model on a flat strip, the boundary
conditions break the symmetries $\delta'$ and $\delta_1$, but preserve the subalgebra
generated by $\delta$ and $\delta_0$. In the untwisted model, this subalgebra
corresponds to the usual B-type supersymmetry considered, for example, in \cite{Hori}.

\subsection{BRST variation of the bulk action and the topological Warner term}

It is not hard to check that the BRST variation of ${\tilde S}_{bulk}$ produces
a boundary term:
\be
\label{deltaSbulk}
\delta {\tilde S}_{bulk} =\delta S_1=\frac{1}{2} \int_{\partial
  \Sigma}{\rho^i\partial_i W}~~.
\ee
The presence of a non-zero right hand side in (\ref{deltaSbulk}) is known as
  the Warner problem \cite{Warner}.

\subsection{$\delta_0$-variation of the bulk action on flat Riemann surfaces}

When considered on a flat Riemann surface, our model has an enlarged symmetry
algebra which was originally described in \cite{Labastida}. In this paper we
shall need only the subalgebra obtained by considering an additional odd
generator $\delta_0$ beyond the BRST generator of (\ref{BRST}). In the
notations of \cite{Labastida}, we have $\delta_0=\{G_0,\}$, where
$G_0=G_z+G_{\bar z}$.  Using the results of \cite{Labastida}, one
finds:
\bea
\label{delta_0}
\delta_0\phi^i=\frac{1}{2}\rho^i_0~~&,&~~\delta_0\phi^{\bar i}=0\nn\\
\delta_0\rho_0^i=0~~&,&~~\delta_0\rho_1^i=-i{\tilde F}^i\\ 
\delta_0\eta^{\bar i}=\partial_0\phi^{\bar i}~~&,&~~\delta_0\theta^{\bar i}=-i\partial_1
\phi^{\bar i}\nn\\ 
\delta_0 {\tilde F}^i=0~~&,&~~\delta_0{\tilde F}^{\bar
i}=\frac{1}{2}(\partial_0\theta^{\bar i}+i\partial_1\eta^{\bar i})~~.\nn
\eea
In this subsection, we are assuming that the worldsheet metric is flat, namely
$g_{\alpha\beta}=\delta_{\alpha\beta}$ in the real coordinates $\sigma^0$ and
$\sigma^1$.

If $\Sigma$ is a cylinder, one finds $\delta_0{\tilde S}_{bulk}=0$. However,
the $\delta_0$ variation of ${\tilde S}_{bulk}$ gives a boundary term when the
model is considered on the strip or on the disk.  Let us take $\Sigma$ to be
infinite the strip given by $(\sigma^0,\sigma^1)\in \R\times [0,\pi]$. We find:
\be \delta_0 {\tilde S}_{bulk}=-\frac{1}{4}\int_{\partial \Sigma}{d\tau
\eta^{\bar i}\partial_{\bar i}{\bar W}}~~.  \ee Here $d\tau=d\sigma^0$ is the
length element along the boundary of $\Sigma$.

\section{Localization formula for correlators on the sphere}
\label{loc_bulk}

Let us consider zero-form bulk observables ${\cal O}$ which are independent
on the auxiliary fields ${\tilde F}^i$ or ${\tilde F}^{\bar i}$.  We are interested in the
sphere correlator of such observables:
\be
\label{correlator}
\langle {\cal O}\rangle_{\rm sphere} =\int{{\cal D}[\phi]{\cal D}[{\tilde F}]
{\cal D}[\eta]{\cal D}[\theta]{\cal D}[\rho]{~e^{-{\tilde S}_{bulk}}{\cal
O}}}~~,
\ee
where we assume that ${\cal O}$ is BRST closed.

In this section, we re-consider the localization formula for such correlators. 
Extending the original argument of \cite{Vafa_LG}, we will extract a
one-parameter family of representations of (\ref{correlator}) as a
finite-dimensional integral. The basic point is a follows. Since the B-model
piece of the Landau-Ginzburg action is BRST exact off-shell, the standard
argument of \cite{Witten_mirror} implies that we can localize on the zero
modes of the associated {\em sigma model}. Thus we shall localize on constant
maps, without requiring that such maps send the worldsheet to the critical
points of $W$. After this reduction, one finds that the Lagrangian density
of the model becomes BRST exact, so the resulting integral representation 
is insensitive to multiplying the Lagrangian density by a prefactor
$\lambda$. Since the former appears multiplied by the worldsheet area,
this prefactor measures the scale of the underlying $S^2$ worldsheet. Thus we
obtain a one-parameter family of localization formulae for our
correlator. Each such "localization picture" allows us to give a geometric
representation of genus zero data, thus providing a geometric model for
the off-shell state space, BRST operator and bulk trace. 

In this approach, the residue representation of \cite{Vafa_LG} is recovered in
the limit $\Re\lambda\rightarrow +\infty$, which forces the point-like image of
the worldsheet to lie on the critical set of $W$. Intuitively, this is the
limit of {\em large} worldsheet areas, the opposite of the
"microscopic" limit $\lambda\rightarrow 0$. 
Varying $\lambda$ allows one to interpolate between these
limits, thus connecting the "sigma-model like" and "residue-like" 
models of genus zero data. 

\subsection{Localization on $B$-model zero modes}

Since ${\tilde S}_B$ is BRST exact, we can replace the bulk
action with:
\be
{\tilde S}_{bulk}=t {\tilde S}_B+ S_W=t\delta V_B+S_W~~,
\ee
where $t$ is a complex parameter with $\Re t >0$ (so that the integral is
well-defined).  BRST invariance of the path integral together with BRST
closure of ${\cal O}$ imply that the resulting correlator is independent of
$t$. This means that (\ref{correlator}) can be computed in the limit
$\Re~t\rightarrow +\infty$, where the integral localizes on the zero-modes of
${\tilde S}_B$. Since $\rho$ has no zero-modes on the sphere, we must
consider configurations for which $\rho^i_\alpha=0$ while $\phi$, $\eta$,
$\theta$ and ${\tilde F}$ are constant on the worldsheet.  For such configurations,
${\tilde S}_B$ reduces to:
\be
{\tilde S}_B|_{\rm zero~modes}=-A G_{i{\bar j}}{\tilde F}^i{\tilde F}^{\bar j}~~,
\ee
with $G_{i{\bar j}}=G_{i{\bar j}}(\phi)$. Here $A$ is the area of the
worldsheet.  The contribution $S_W$ becomes:
\be
S_W|_{\rm zero~modes}=-\frac{i}{2}A\left[ D_{\bar i} \partial_{\bar j}{\bar W}
\chi^{\bar i}{\bar \chi}^{\bar j}- (\partial_{\bar i}{\bar W}){\tilde F}^{\bar
i}+ (\partial_i W){\tilde F}^i\right]~~.
\ee
Combining these expressions, we find the zero-mode reduction of the worldsheet
action:
\be
{\tilde S}_{0}:={\tilde S}_{bulk}|_{\rm zero~modes}=-A G_{i{\bar j}}{\tilde
F}^i{\tilde F}^{\bar j}- \frac{i}{2}A\left[ \frac{1}{2}D_{\bar i}
\partial_{\bar j}{\bar W} \theta^{\bar i}\eta^{\bar j}- (\partial_{\bar
i}{\bar W}){\tilde F}^{\bar i}+ (\partial_i W){\tilde F}^i\right]~~,
\ee
where we wrote $\chi$ and ${\bar \chi}$ in terms of $\eta$ and $\theta$.
The correlator (\ref{correlator}) reduces to an ordinary integral over the
zero-modes $\phi^i, \phi^{\bar i}$, ${\tilde F}^i, {\tilde F}^{\bar i}$ and
$\eta^{\bar i},\theta_i$:
\be
\label{correlator_0}
\langle {\cal O}\rangle_{\rm sphere} =\int{d\phi d{\tilde F} d\eta d\theta
{~e^{-{\tilde S}_{0}}{\cal O}}}~~.
\ee
On zero modes, the BRST generator (\ref{BRST}) takes the form:
\bea
\label{BRST_0}
\delta \phi^i=0~~&,&~~\delta \phi^{\bar i}=\eta^{\bar i}\nn\\ \delta
\eta^{\bar i}=0~~&,&~~\delta \theta^{\bar i}=2{\tilde F}^{\bar i}+\Gamma^{\bar
i}_{{\bar j}{\bar k}}\theta^{\bar j}\eta^{\bar k}~~\\ \delta {\tilde
F}^i=0~~&,&~~\delta {\tilde F}^{\bar i}=\Gamma^{\bar i}_{{\bar j}{\bar
k}}{\tilde F}^{\bar j}\eta^{\bar k}~~.\nn
\eea
In particular, we have:
\be
\label{tildeS_B_0_ex}
G_{i{\bar j}}{\tilde F}^i{\tilde F}^{\bar j}=\frac{1}{2}\delta [G_{i{\bar j}}{\tilde
  F}^i\theta^{\bar j}]~~,
\ee
which is the zero-mode remnant of equation (\ref{exactS_B}).  One can also
check directly that the reduced action is BRST-closed.

The integral over ${\tilde F}$ can be cast into Gaussian form through the change
of variables:
\be
{\tilde F}^i={\hat F}^i+\frac{i}{2}G^{i{\bar j}}\partial_{\bar j}{\bar W}~~,~~
{\tilde F}^{\bar i}={\hat F}^{\bar i}-\frac{i}{2}G^{{\bar i}j}\partial_j W~~.
\ee
Then the reduced action becomes: 
\be
{\tilde S}_{0}=-AG_{i{\bar j}}{\hat F}^i{\hat F}^{\bar j}+
\frac{1}{4}A \left[
-iD_{\bar i} \partial_{\bar j}{\bar W} \theta^{\bar
      i}\eta^{\bar j}+G^{i{\bar j}}(\partial_i W)(\partial_{\bar j}{\bar W})\right]~~.
\ee
Integrating over ${\hat F}$, we find:
\be
\label{correlator_0intmd}
\langle {\cal O}\rangle_{\rm sphere}=N \int{d\phi d\eta d\theta
e^{-\frac{A}{4} {\tilde L}_0}{\cal O}}~~,
\ee
where:
\be
\label{tildeL_0}
{\tilde L}_0:=-i D_{\bar i} \partial_{\bar j}{\bar W} \theta^{\bar
      i}\eta^{\bar j}+G^{i{\bar j}}(\partial_i W)(\partial_{\bar j}{\bar
      W})~~
\ee
plays the role of zero-mode Lagrange density. The prefactor in 
(\ref{correlator_0intmd})  has the form:
\be
N=\frac{(2\pi)^n}{ A^n \det(G_{i{\bar j}})}~~,
\ee
where $n$ is the complex dimension of the target space $X$. Since we
integrated out the fields ${\hat F}$, the BRST generator on
zero-modes reduces to:
\bea
\label{BRST_os_0}
\delta \phi^i=0~~&,&~~\delta \phi^{\bar i}=\eta^{\bar i}\nn\\
\delta \eta^{\bar i}=0~~&,&~~\delta \theta^{\bar
  i}=-iG^{{\bar i}j}\partial_jW+ \Gamma^{\bar i}_{{\bar j}{\bar
  k}}\theta^{\bar j}\eta^{\bar k}~~,
\eea
which is obtained from  (\ref{BRST_0}) by imposing the equations of motion:
\be
{\tilde F}^i=\frac{i}{2}G^{i{\bar j}}\partial_{\bar j}{\bar W}~~,~~
{\tilde  F}^{\bar i}=-\frac{i}{2}G^{{\bar i}j}\partial_jW~~.
\ee
For later reference, notice that the last relation in (\ref{BRST_os_0}) is equivalent with:
\be
\label{delta_theta_os}
\delta \theta_i=-i\partial_i W~~.
\ee

Using this form of the BRST transformations, one finds that the zero-mode
Lagrange density (\ref{tildeL_0}) is BRST exact:
\be
\label{exact_tildeL_0}
{\tilde L}_0=\delta {\tilde v}_0~~,
\ee
where:
\be
\label{tildev_0}
{\tilde v}_0=i\theta^{\bar i}\partial_{\bar i}{\bar W}~~.
\ee
Thus we can replace  (\ref{correlator_0intmd}) by:
\be
\label{correlator_final}
\langle {\cal O}\rangle_{\rm sphere} 
=N\int{d\phi d\eta d\theta {~e^{-\lambda {\tilde L}_0}{\cal O}}}~~,
\ee
where $\lambda$ is a complex parameter with positive real part. The integral
(\ref{correlator_final}) is independent of its value.

\subsection{The space of bulk observables and its cohomology}
\label{bulk_observables}

The observables of interest have the form:
\be
\label{bulk_obs}
{\cal O}_\omega(\sigma):=\omega^{j_1\dots j_q}_{{\bar i}_1\dots {\bar
    i}_p}(\phi(\sigma))\eta^{{\bar
    i}_1}(\sigma) \dots \eta^{{\bar i}_p}(\sigma)
\theta_{j_1}(\sigma) \dots \theta_{j_q}(\sigma)~~,
\ee
where $\omega:=\omega^{j_1\dots j_q}_{{\bar i}_1\dots {\bar i}_p}dz^{{\bar
  i}_1}\wedge \dots \wedge dz^{{\bar i}_p}\wedge \partial_{j_1}\wedge \dots \wedge \partial_{j_q}$
is a section of the bundle 
$\Lambda^p{{\bar T}^*X}\wedge \Lambda^q TX$. After reduction to B-model 
zero modes, we are left with:
\be
{\cal O}_\omega=\omega^{j_1\dots j_q}_{{\bar i}_1\dots {\bar
    i}_p}(\phi)\eta^{{\bar
    i}_1} \dots \eta^{{\bar i}_p}
\theta_{j_1}\dots \theta_{j_q}~~,
\ee
which can be identified with the polyvector-valued form $\omega$ upon setting:
\be
\label{ids_geom}
\eta^{\bar j}\equiv dz^{\bar j}~~,~~\theta_j\equiv \partial_j~~. 
\ee

The reduced BRST operator (\ref{BRST_os_0})  becomes:
\be
\label{delta_geom}
\delta\equiv {\bar \partial}+i_{\partial W}
\ee
where $i_{\partial_W}$ is the odd derivation of $\Lambda^*TX$ uniquely determined
by the conditions:
\be
i_{\partial W}(\partial_j)=-i\partial W(\partial_j)=-i\partial_j W~~.
\ee
Thus the cohomology of the differential superalgebra 
$({\cal H}, \delta)$, where ${\cal H}:=\Gamma(\Lambda^*{\bar T}^*X\wedge \Lambda^*TX)$, models
the algebra of bulk observables. 
The obvious  relations:
\be
{\bar \partial}^2=(i_{\partial W})^2={\bar
  \partial}i_{\partial_W}+i_{\partial_W}{\bar \partial}=0~~
\ee
show that $({\cal H},\delta)$ is a bicomplex. Hence the BRST cohomology is
computed by a spectral sequence $E_*$ whose second term equals:
\be
\label{E}
E_2:=H_{i_{\partial_W}}(H_{\bar \partial}({\cal H}))~~.
\ee
Since the target space is non-compact, we must of course specify a growth condition at
infinity. We shall take ${\cal H}$ to consist of those sections of 
the bundle $\Lambda^*{\bar T}^*X\wedge \Lambda^*TX$ whose coefficients have at most
polynomial growth. When the spectral sequence collapses to its second term,
the BRST cohomology reduces to (\ref{E}). A standard example is the case
$X=\C^n$, with $W$ a polynomial function of $n$ variables. 
Then the $\bar{\partial}$-Poincare Lemma implies 
that $H_{\bar \partial}({\cal H})$ coincides with the space
$\Gamma_{poly}(\Lambda^*TX)$ of polyvector fields with polynomial
coefficients.  In this case, the BRST cohomology reduces to the Jacobi ring
$\C[x_1\dots x_n]/\langle \partial_1 W\dots \partial_n W\rangle$, thereby
recovering a well-known result.

\subsection{The geometric model}
\label{bulk_geom}

Let us translate (\ref{correlator_final}) into
classical mathematical language. Using (\ref{ids_geom}), we find:
\be
\label{tildeL_0form_0}
{\tilde L}_0\equiv i{\bar H}^{i}_{~{\bar j}}
dz^{\bar j}\wedge \partial_{i}+
G^{i{\bar j}}(\partial_iW)(\partial_{\bar j} {\bar W}) \in {\cal H}~~.
\ee
Here $H^i_{~{\bar j}}:=G^{i{\bar k}}{\bar H}_{{\bar k}{\bar j}}$, where 
$H_{ij}:=D_i\partial_j W$ is the Hessian of $W$. Consider the Hessian
operator:
\be
H=H^{\bar i}_{~j}dz^j\otimes \partial_{\bar i}\in Hom(TX,{\bar T}X)=T^*X\otimes
{\bar T}X~~,
\ee
whose complex conjugate has the form:
\be
{\bar H}={\bar H}^i_{~{\bar j}}dz^{\bar j}\otimes \partial_i\in Hom({\bar
  T}X,TX)={\bar T}^*X\otimes TX~~.
\ee
The quantity
${\bar H}_a:={\bar H}^i_{~{\bar j}}dz^{\bar j}\wedge \partial_i\in {\bar T}^*X\wedge TX$ 
appearing in (\ref{tildeL_0form_0}) is the antisymmetric part of
${\bar H}$.  On the other hand, the second term of (\ref{tildeL_0form_0}) is the norm of the
differential form $\partial W=\partial_iW dz^{i}$. This gives the
coordinate-independent version of (\ref{tildeL_0form_0}):
\be
{\tilde L}_0=i{\bar H}_{a}+||\partial W||^2~~.
\ee
Also note the representation:
\be
{\tilde v}_0=iG^{i{\bar j}}\partial_i\wedge \partial_{\bar j} {\bar W}~~.
\ee
It is now easy to see that (\ref{correlator_final}) becomes:
\be
\label{correlator_geom}
\Tr\omega:=\langle {\cal O}_\omega\rangle_{\rm sphere}=N\int_{X}{\Omega\wedge \left[
\Omega \lrcorner \left(e^{-\lambda {\tilde L}_0}\wedge \omega\right)\right]}~~,
\ee
where $\lrcorner$ denotes the total contraction of a form with a
polyvector. The linear functional $\Tr$ realizes the bulk trace of \cite{CIL1}.

\paragraph{Observation}
The integral representation (\ref{correlator_geom}) allows us 
to give another (and completely rigorous) 
proof of $\lambda$-independence for $\delta
\omega=0$, with the assumption $\Re \lambda>0$. 
For this, we have to show that the $\lambda$-derivative of
(\ref{correlator_geom}) vanishes.
Since ${\tilde L}_0=\delta {\tilde v}_0$ and $\delta \omega=0$, this
derivative takes the form:
\be
\label{der_geom}
\frac{d}{d\lambda}\Tr \omega =
-\lambda N\int_{X}{\Omega\wedge \left[\Omega \lrcorner 
\delta \left(e^{-\lambda {\tilde L}_0}\wedge {\tilde v}_0\wedge \omega\right)\right]}~~.
\ee
Thus it suffices to show that 
$\int_{X}{\Omega\wedge \left[\Omega \lrcorner \delta \alpha\right]}$
vanishes for any $\alpha\in {\cal H}$ which decays exponentially at infinity
on $X$ (the exponential  decay for $\alpha=e^{-\lambda {\tilde L}_0}{\tilde
  v}_0\wedge \omega$
in (\ref{der_geom}) is due to the second term in (\ref{tildeL_0form_0})). 
Notice further that $\int_{X}{\Omega\wedge \left[\Omega
  \lrcorner \delta \alpha\right]}$
vanishes for degree reasons unless $\delta \alpha\in \Gamma(\Lambda^n{\bar T}^*X\wedge \Lambda^n
TX)$. Hence it is enough to show vanishing of $\int_{X}{\Omega\wedge \left[
\Omega \lrcorner \delta \alpha\right]}$ for an exponentially decaying $\alpha$
such that
$\delta \alpha\in \Gamma(\Lambda^n{\bar T}^*X\wedge \Lambda^n TX)$. In this
case, we obviously have $\delta\alpha={\bar \partial}\beta$ for some 
exponentially decaying $\beta\in \Gamma(\Lambda^{n-1}{\bar T}^*X\wedge
\Lambda^n TX)$ (this follows by noticing that the
image of $i_{\partial W}$ has vanishing intersection with the subspace 
$\Gamma(\Lambda^n{\bar T}^*X\wedge \Lambda^n TX)$). Therefore, we only need to show
that $\int_{X}{\Omega\wedge \left[\Omega \lrcorner {\bar \partial} \beta\right]}$
vanishes. This last fact follows from 
$\Omega\wedge \left[\Omega \lrcorner {\bar \partial}
  \beta\right]={\bar \partial}\left(\Omega\wedge \left[\Omega
    \lrcorner \beta\right]\right)$, since the boundary term
vanishes due to the exponential decay of $\beta$. 
The assumption $\Re\lambda>0$ is crucial,
since otherwise we cannot rely on exponential decay to conclude that the
boundary term vanishes.

\subsection{Localization pictures and homotopy flows}
\label{bulk_flow}

Expression (\ref{correlator_geom}) admits the following interpretation.
Consider the one-parameter semigroup of operators $U(\lambda)$ acting on
${\cal H}$ through wedge multiplication by $e^{-\lambda{\tilde L}_0}$:
\be
U(\lambda)\omega:=e^{-\lambda{\tilde L}_0}\wedge \omega~~{\rm~for~all}~~\omega\in {\cal H}~~.
\ee
The semigroup is defined on the half-plane $\Delta:=\{\lambda\in
\C|\Re\lambda>0\}$, so that $U(\lambda)$ maps ${\cal H}$ into a subspace of
itself. Then (\ref{correlator_geom}) takes the form: 
\be
\label{Tr_flow}
\Tr\omega:=\Tr^B(U(\lambda)\omega)~~,
\ee
where $\Tr^B$ is the bulk trace of the B-twisted sigma model: 
\be
\Tr^B\omega:=N\int_{X}{\Omega\wedge \left(
\Omega \lrcorner \omega\right)}~~.
\ee

Since ${\tilde L}_0$ is BRST closed (${\tilde L}_0=\delta {\tilde v}_0$), each
operator $U(\lambda)$ is homotopy equivalent with the identity in the complex $({\cal H},
\delta)$:
\be
U(\lambda)=1+[\delta, W_\lambda]~~,
\ee
for some operator $W_\lambda$. In particular, $U(\lambda)$ is an endomorphism of our
complex, i.e. the following relation holds: 
\be
\label{endomorphism}
U(\lambda)\circ \delta=\delta \circ U(\lambda)~~. 
\ee
Such a semigroup will be called a {\em homotopy
flow}. It is clear that each $U(\lambda)$ is a quasi-isomorphism, i.e. 
induces an automorphism $U_*(\lambda)$ on the BRST cohomology $H_\delta({\cal
  H})$. Following relation (\ref{Tr_flow}), we define the {\em localization
  picture $\lambda$} by associating $\omega_\lambda:=U(\lambda)(\omega)\in
{\cal H}$ to each $\omega\in {\cal H}$ (then $\omega_\lambda$ is the representative of the "state"
$\omega$ in the picture $\lambda$). The representatives of this
picture belong to the subspace ${\cal H}_\lambda:=U(\lambda)({\cal
  H})\subset {\cal H}$. As in quantum mechanics, we have a representative for
any operator $T\in End({\cal H})$ in the localization picture $\lambda$:
\be
T_\lambda:=U(\lambda)\circ T \circ U(-\lambda)\in End({\cal H}_\lambda)~~,
\ee
where $\Re\lambda>0$ and $U(-\lambda)$ is defined as an operator from ${\cal
  H}_\lambda$ to ${\cal H}$. Relation (\ref{endomorphism})
shows that the BRST operator is "picture-independent" in the following sense: 
\be
Q_\lambda=Q|_{{\cal H}_\lambda}~~,
\ee
where in the right hand side we restrict both the domain and image 
to ${\cal H}_\lambda$. Relation (\ref{Tr_flow}) becomes:
\be
\Tr\omega=\Tr^B\omega_\lambda~~. 
\ee

For $\lambda=W=0$, we have $U(0)=Id_{\cal H}$ and we recover the familiar data
of the B-twisted sigma model. Namely, ${\cal H}$ provides a geometric model for the
off-shell state space, the Dolbeault operator ${\bar \partial}$ models the "localized
BRST operator" and $\Tr^B$ models the bulk trace of \cite{CIL1}. 
Turning on the Landau-Ginzburg superpotential $W$ and
performing localization as above with "worldsheet area" $\lambda$  leads to a
geometric model given by the triplet $({\cal H}, \delta, \Tr)$. This is
related to the triplet describing the B-twisted sigma model by the modification
$\delta={\bar \partial}+i_{\partial W}$ of the BRST operator, followed by the
homotopy flow $U(\lambda)$. 

Because varying $\lambda$ along the real axis amounts to changing the area of the
worldsheet, the operators $U(\lambda)$ implement a sort of
"renormalization group flow" connecting the point-like (UV) limit $\lambda=0$
with the large area (IR limit) $\lambda=+\infty$. Since the model is
topological, such a flow "does nothing" at the level of BRST cohomology, but
acts non-trivially off-shell.

\subsection{The residue formula for sphere correlators}

Since the integral (\ref{correlator_final}) is independent of $\lambda$, we
can compute its value for $\Re \lambda\rightarrow +\infty$ (more specifically,
we shall take $\lambda\rightarrow +\infty$ with $\lambda \in \R$). In this limit,
the second term in (\ref{tildeL_0}) forces the integral to localize on the
critical points of $W$, and
the Gaussian approximation  around these points 
becomes exact. For simplicity, we shall assume that the critical points of $W$
are isolated (the general case can be incorporated by a continuity argument). 
For simplicity, we shall also assume that the spectral sequence of Subsection
\ref{bulk_observables} collapses to its second term. 

Taking $\lambda\rightarrow +\infty$ with $\lambda \in \R$, we find that the correlator 
(\ref{correlator_final}) vanishes unless $\omega=f$ with $f$ a complex-valued 
function defined on $X$ \footnote{For this, notice that the bosonic Gaussian integral over
  fluctuations of $\phi$ around
  each critical point of $W$ produces a factor which is weighted by
  $\frac{1}{\lambda^n}$. Thus the fermionic Gaussian integral over $\theta$
  and $\eta$ must produce $n$ powers of $\lambda$ if one is to obtain a
  non-vanishing result in the limit $\lambda\rightarrow \infty$. This
  obviously requires that ${\cal O}_\omega$ contain no $\eta$'s or $\theta$'s, 
  so that the highest ($n$-th order term) in the expansion of $e^{+i\lambda D_{\bar i}
    \partial_{\bar j}{\bar W} \theta^{\bar i}\eta^{\bar j}}$ survives when
  performing the integral over $\eta$ and $\theta$. }. In this case, we obtain:
\bea &&\langle {\cal O}_f \rangle_{\rm sphere}=\nn\\
&=&\lim_{\lambda\rightarrow +\infty} \left(N\sum_{p\in {\rm Crit}
W}{\left[n!^2 (i\lambda)^n (-1)^{n(n-1)/2}\det ({\bar H}_{{\bar i}{\bar
j}})\right]\left[\frac{(2\pi)^n}{\lambda^n \det (H_{ij}) \det ({\bar H}_{{\bar
i}{\bar j}})}\det (G_{i{\bar j}})\right]}f|_{p}+O(1/\lambda)\right)\nn\\
&=&n!^2 (2\pi)^n (-1)^{n(n+1)/2}N\det (G_{i{\bar j}})\sum_{p\in {\rm Crit} W}{
\frac{1}{\det (H_{ij}(p))} f(p)}~~.  \eea
Since $\sum_{p\in {\rm Crit} W}{\frac{1}{\det (H_{ij}(p))} f(p)}
\propto \int_{X}{\frac{f(z)dz^1\wedge\dots \wedge dz^n }{\partial_1W\dots
  \partial_n W}}$ by residue theory \cite{Griffiths}, 
one recovers the following generalization of the well-known result of \cite{Vafa_LG}:
\bea
\langle {\cal O}_\omega \rangle_{\rm sphere}&=&0~~{\rm unless}~~\omega=f \\
\langle {\cal O}_f\rangle_{\rm sphere}&=& C
\int_{X}{\Omega \frac{f(z) }{\partial_1W\dots  \partial_n W}}~~.
\eea
Here $C$ is an uninteresting normalization constant.

\section{The boundary coupling}
\label{boundary_coupling}

In this section we discuss the boundary coupling of our models. The
construction is based on \cite{coupling}, with a certain modification which
will prove useful later on. After recalling the basics of 
superconnections, we construct the coupling in the form of \cite{coupling},
with the addition of a term which insures $\delta_0$-invariance on a flat
strip. While this does not affect the target space equations of motion,
it will help us make contact with previous work on the subject. We also give
the target space reflection of the $\delta_0$-invariance constraint.

\subsection{Mathematical preparations}

Consider a complex superbundle $E=E_+\oplus E_-$ over $X$, and a
superconnection \cite{Quillen} ${\cal B}$ on $E$. We let $r_\pm:=\rk E_\pm$. 
The bundle of endomorphisms $End(E)$ is endowed with the 
natural $\Z_2$ grading, with even and odd components:
\bea
End_+(E)&:=&~~~~~End(E_+)\oplus End(E_-)\\
End_-(E)&:=& Hom(E_+,E_-)\oplus Hom(E_-,E_+)~~.
\eea
The superconnection ${\cal B}$  can be viewed 
as a section of $[{\cal T}^*X\otimes End_+(E)]\oplus End_-(E)$. 
In a local frame of $E$ compatible with the grading, this is a matrix:
\be
{\cal B}=\left[\ba{cc} A^{(+)}& F\\G & A^{(-)}\ea\right]
\ee
whose diagonal entries $A^{(\pm)}$ are connection one-forms on $E_\pm$, while 
$F,G$ are elements of  $Hom(E_-,E_+)$ and $Hom(E_+,E_-)$. We 
require that the superconnection has type $(0,\leq 1)$, i.e. the one-forms
$A^{(\pm)}$ belong to $\Omega^{(0,1)}(End(E_\pm))$. The morphism $F$ should not be
confused with the curvature form used below.

When endowed with the ordinary composition of morphisms, the space of sections
$\Gamma(End(E))$ becomes an associative superalgebra. The space
${\cal H}_b:=\Omega^{(0,*)}(End(E))$ also carries an associative superalgebra structure,
which is induced from $(\Omega^{(0,*)}(X),\wedge)$ and 
$(\Gamma(End(E)),\circ)$ via the tensor product decomposition:
\be
\Omega^{(0,*)}(End(E))=\Omega^{(0,*)}(X)\otimes_{\Omega^{(0,0)}(X)} \Gamma(End(E))~~.
\ee
For decomposable
elements $u=\omega\otimes f$ and $v=\eta\otimes g$, with
homogeneous  
$\omega,\eta$ and $f,g$, the associative product on ${\cal H}_b$ takes the form:
\be
uv=(-1)^{ \deg f~\rk \eta}(\omega\wedge \eta)\otimes (f\circ g)~~,
\ee
where $\deg$ denotes the grading of the superalgebra $End(E)$:
\be
\deg(f)=0\in \Z_2~~{\rm~if~}f\in End_+(E)~~,~~\deg(f)=1\in \Z_2~~{\rm~if~}f\in End_-(E)~~.
\ee
The total degree on ${\cal H}_b$ is given by:
\be
|\omega\otimes f|=\rk \omega +\deg f~~(mod~2)~~.
\ee
We also recall the supertrace on $End(E)$:
\be
\label{str}
\str(f)=\tr f_{++}-\tr f_{--}~~,
\ee
where $f=\left[\ba{cc}f_{++}&f_{-+}\\f_{+-}&f_{--}\ea\right]$ is an
endomorphism of $E$ with components $f_{\alpha\beta}\in Hom(E_\alpha,E_\beta)$
for $\alpha,\beta=+,-$. This has the property:
\be
\label{str_cyc}
\str(f\circ g)=(-1)^{\deg f \deg g}\str(g\circ f)~~
\ee
for homogeneous elements $f,g$. 

The twisted Dolbeault operator:
\be
\label{sc}
{\bar {\cal D}}={\overline \partial} +{\cal B}=\left[\ba{cc} 
{\bar \partial}+A^{(+)}& F\\G &{\bar \partial}+A^{(-)}\ea\right]
\ee
induces an odd derivation ${\bar \partial}+[{\cal B},\cdot]$ 
of the superalgebra ${\cal H}_b$, where
$[u,v]:=uv-(-1)^{|u||v|}vu$ is the supercommutator. 

The $(0,\leq 2)$ part of the superconnection's curvature has the form:
\be
\label{curvature}
{\cal F}^{(0,\leq 2)}={\bar {\cal D}}^2={\overline \partial}
{\cal B}+\frac{1}{2}[{\cal B}, {\cal B}]={\overline \partial}
{\cal B}+{\cal B}{\cal B}=
\left[\ba{cc} F^{(+)}_{(0,2)}+FG& {\bar \nabla} F\\{\bar \nabla} G &F^{(-)}_{(0,2)}+GF\ea\right]
\ee
where $F^{(\pm)}_{(0,2)}$ are the $(0,2)$ pieces of the curvature forms
$F^{(\pm)}$ of $A^{(\pm)}$ and:
\bea
{\bar \nabla}F&=&{\bar \partial}F+
A^{(+)}F+FA^{(-)}={\bar \partial}F+A^{(+)}\circ F-F\circ A^{(-)}~~\nn\\
{\bar \nabla}G&=&{\bar \partial}G+A^{(-)}G+G A^{(+)}
={\bar \partial} G+A^{(-)}\circ G-G\circ A^{(+)}~~.
\eea

We will use the the notations:
\be
\label{AD}
A:=A^{(+)}\oplus A^{(-)}=\left[\ba{cc} A^{(+)}& 0\\0 & A^{(-)}\ea\right]~~,~~
D:=\left[\ba{cc} 0& F\\G & 0\ea\right]
\ee
for the diagonal and off-diagonal parts of ${\cal B}$. Then $A$ is an
connection one-form on $E$ (compatible with the grading), while
$D$ is an odd endomorphism. We have ${\cal B}=A+D$ and:
\be
\label{02curvature}
{\cal F}^{(0,\leq 2)}=F^{(0,2)}+{\bar \nabla}_A D +D^2~~.
\ee
Here $F^{(0,2)}=F^{(+)}_{(0,2)}+F^{(-)}_{(0,2)}$ is the $(0,2)$ part of the curvature of $A$ and 
${\bar \nabla}_A={\bar \partial}+[A,\cdot]$ is the Dolbeault operator twisted by $A$.  

\subsection{The boundary coupling}

Following \cite{coupling}, we define the partition function on a bordered and
oriented Riemann surface $\Sigma$ by:
\be
\label{Z}
Z:=\int{{\cal D}[\phi]{\cal D}[{\tilde F}]{\cal D}[\theta]{\cal D}[\rho]{\cal D}[\eta]
  e^{-{\tilde S}_{bulk}} {\cal U}_1\dots {\cal U}_h}~~,
\ee
where $h$ is the number of holes and the factors ${\cal U}_a$ have the form:
\be
\label{calU}
{\cal U}_a:=\Str Pe^{-\oint_{C_a}{d\tau_a M}}~~.
\ee
We are assuming that the boundary of $\Sigma$ is a disjoint union of 
smooth circles $C_a$, associated with holes labeled by $a$. 
The symbol $\Str$ denotes the supertrace on $GL(r_+|r_-)$, while $d\tau_a$ 
stands for the length element along $C_a$ induced by the metric on the
interior of $\Sigma$. The quantity $M$ is given by:
\be
\label{M}
M=\left[\ba{cc} 
{\hat {\cal A}}^{(+)}+\frac{i}{2}(FF^\dagger+G^\dagger G) &
\frac{1}{2}\rho^i_0 \nabla_i F+\frac{i}{2}\eta^{\bar i}\nabla_{\bar i}G^\dagger
\\
\frac{1}{2} \rho^i_0 \nabla_i G +\frac{i}{2}\eta^{\bar i}\nabla_{\bar
  i}F^\dagger &
{\hat {\cal A}}^{(-)} +\frac{i}{2}(F^\dagger F+GG^\dagger)
\ea\right]~~.
\ee
Here $\rho_0^i d\tau_a$ is the pull-back of $\rho^i$ to $C_a$ and:
\be
\label{calApm} 
{\hat {\cal A}}^{(\pm)}:=A_{\bar i}^{(\pm)}{\dot \phi}^{\bar i} +\frac{1}{2}\eta^{\bar i}  
F^{(\pm)}_{{\bar i} j}\rho_0^j
\ee
are connections on the bundles ${\cal E}_\pm$ obtained by pulling back
$E_\pm$ to the boundary of $\Sigma$. The dot in (\ref{calApm}) stands for the
derivative $\frac{d}{d\tau_a}$. Notice that $\nabla_i F=\partial_i F$ and
$\nabla_i G=\partial_i G$ since $A$ is a $(0,1)$-connection.

We have:
\be
\label{M_dec}
M={\hat {\cal A}}+\Delta+K
\ee
where: 
\be
\Delta:=\frac{1}{2}\rho_0^i\partial_i D~~,
\ee
\be
K:=\frac{i}{2}\left(\eta^{\bar i}\nabla_{\bar i}D^\dagger +[D,D^\dagger]_+ \right)
\ee
and:
\be
{\hat {\cal A}}={\dot \phi}^{\bar i}A_{\bar i}+\frac{1}{2}F_{{\bar
    i}j}\eta^{\bar i}\rho^j_0~~.
\ee
Here  $A$ is the direct sum connection on $End(E)$ introduced in
(\ref{AD}). The first two terms in (\ref{M_dec}) agree with \cite{coupling},
while the last term $K$ is added for comparison with \cite{Kap2}. As we shall
see below, this term preserves BRST-invariance of the partition function
(which is already preserved by the sum of the first two terms
\cite{coupling}). As for the open B-model,
adding $K$ insures invariance of the boundary coupling with respect to 
the second generator $\delta_0$ of the $N=2$ topological algebra, thereby
fixing an ambiguity familiar form Hodge theory\footnote{
The symmetry generators $\delta$ and $\delta_0$ can be viewed as analogues of
the operators ${\bar \partial}$ and ${\bar \partial}^\dagger$ of Hodge theory, as already pointed out in \cite{Witten_mirror}
in the context of twisted B-models. The boundary coupling of \cite{coupling}
is chosen to preserve BRST invariance of the partition function. This is
ambiguous up to addition of 'exact' terms, an ambiguity which we can
fix by requiring $\delta_0$-invariance of the partition function. }. 
This modification has minor effects which can be safely
ignored for most purposes 
\footnote{As we shall see in the next section, the extra-term in the boundary
  coupling can be used to 
introduce a parameter $\mu$ characterizing boundary localization pictures. 
For most practical purposes, this parameter can be set to zero, which amounts
to neglecting the last term in (\ref{M_dec}). In
particular, one must set $\mu$ to zero in order to recover the trace
formula of \cite{Kap2}. It is the {\em bulk} parameter $\lambda$ which must be
taken to infinity in order to recover the proposal of \cite{Kap2}.}. 

\subsection{The target space equations of motion}

To insure BRST invariance of the partition function 
(\ref{Z}), we must choose the background superconnection
${\cal B}$ such that:
\be
\label{invar}
\delta {\cal U}_a = \frac{1}{2}\left[\int_{C_a}{d\tau
    \rho^i_0\partial_i W} \right] {\cal U}_a~~.
\ee
In this paper, we also require $\delta_0$-invariance of the partition function on the flat
strip:
\be
\label{invar_0}
\delta_0 {\cal U}_a = -\frac{1}{4}\left[\int_{C_a}{d\tau
    \eta^{\bar i}\partial_i {\bar W}} \right] {\cal U}_a~~.
\ee
It is not hard to check the relations:
\be
\label{Qhol}
\delta {\cal U}_a=-\Str \left[~I_a(\delta M)
  Pe^{-\oint_{C_a}{d\tau_a M}}\right]
\ee
where:
\bea
I_a(\delta M)&=&\oint_{C_a}d\tau_a 
  U_a^{-1}\Big(F_{{\bar i}{\bar j}}\eta^{\bar i}{\dot \phi}^{\bar
      j}-\frac{1}{4} \nabla_k F_{{\bar i}{\bar j}}\eta^{\bar i}\eta^{\bar
      j}\rho_0^k-{\dot \phi}^{\bar
      i}\nabla_{\bar
      i}D-\frac{1}{2}\rho^i_0\nabla_i(D^2)+\frac{1}{2}\eta^{\bar
      i}\rho^j_0\nabla_j\nabla_{\bar i}D+\nn\\
&&+\frac{1}{2}\eta^{\bar i}\eta^{\bar j}[F_{{\bar i}{\bar
  j}},D^\dagger]+\eta^{\bar i}[\nabla_{\bar i}D,D^\dagger]+[D^2,D^\dagger]
\Big)U_a~,~~~~\nn
\eea
and:
\be
\label{delta_0hol}
\delta_0 {\cal U}_a=-\Str \left[~I_a(\delta_0 M)
  Pe^{-\oint_{C_a}{d\tau_a M}}\right]
\ee
where:
\bea
I_a(\delta_0 M)&=&\frac{1}{4}\oint_{C_a}{d\tau_a 
U_a^{-1}\left(\eta^{\bar i}\nabla_{\bar i}(D^\dagger)^2+[D,(D^\dagger)^2]\right)U_a}\nn\\
&+&\frac{i}{2}\oint_{C_a}{d\tau_a 
U_a^{-1}\left( -{\dot \phi}^i \nabla_i D^\dagger +\frac{1}{2}\eta^{\bar i}\nabla_{\bar
  i}\nabla_j D^\dagger \rho_0^j+\frac{1}{2}\rho_0^i [D,\nabla_i D^\dagger]
  \right) U_a}~~.~~~~\nn
\eea
Here $U_a(\tau_a)\in GL(r_+|r_-)$ is a certain invertible
  operator\footnote{This should not be confused with the homotopy flow of
  Subsection \ref{bulk_flow} !}) which plays
the role of `parallel transport' defined by $M$ along $C_a$ (see
\cite{coupling} for details). Namely $U_a(\tau_a)=U_a(\tau_a, 0)$, where:  
\be
\label{parr_tr}
U_a(\tau_2,\tau_1):=Pe^{-\int_{\tau_1}^{\tau_2} M(\tau)d\tau}~~
\ee
if $\tau_2>\tau_1$. The origin of the proper length coordinate $\tau_a$ 
along $C_a$ is chosen arbitrarily, while the orientation on $C_a$ is
compatible with that of $\Sigma$. 
The quantities   $F_{{\bar i}{\bar j}}$ etc. are the 
$(0,2)$-components of the curvature of the direct sum connection $A$ introduced in (\ref{AD}).
Notice the relations: 
\be
{\cal U}_a=\Str H_a(\tau)~~
\ee
where: 
\be
\label{holonomy}
H_a(\tau)=U(\tau+l_a,\tau)
\ee 
are the "superholonomy operators" (here $l_a$ the length of $C_a$).

Relations (\ref{invar}, \ref{invar_0}) and (\ref{Qhol},\ref{delta_0hol}) 
show that the BRST and $\delta_0$-invariance conditions amount to:
\bea
\label{invariance_eqs}
&&F_{{\bar i}{\bar j}}=0\\
&&\nabla_{\bar i} D=0\\
&&\nabla_i(D^2)=\partial_iW\\
&&[D^\dagger,D^2]=0~~.
\eea
The first relation says that $A$ is integrable, so it defines a complex
structure on the bundle $E$. The second condition means that $D\in End(E)$ is
holomorphic with respect to this complex structure. The third equation 
requires $D^2=c+W {\rm id}_E$, with $c$ a covariantly-constant endomorphism. Comparing with
(\ref{02curvature}), we find that these first three conditions are equivalent with:
\bea
\label{eom}
{\cal F}^{(0,\leq 2)}=c+W{\rm id}_E\Longleftrightarrow {\cal {\bar D}}^2=c+W{\rm id}_E~~.
\eea
This is the target space equation of motion for our open string background
\cite{coupling}. Notice that (\ref{eom}) admit solutions only when
$r_+=r_-$.

For backgrounds satisfying the equation of motion, 
the last condition in (\ref{invariance_eqs}) reads:
\be
[D^\dagger,D^2]=0 \Longleftrightarrow [D^\dagger,c]=0~~. 
\ee
This can be viewed as a partial "gauge-fixing" constraint,
which is fulfilled, for example, if one takes $c$ to be proportional to the
identity endomorphism (in which case the proportionality constant can be absorbed into $W$). 
For simplicity, we shall take $c=0$ for the remainder of this paper. 

\section{Boundary observables and correlators}

As we  shall see in Section \ref{loc_boundary}, 
the boundary conditions derived from the partition function (\ref{Z}) constrain $\theta$ in terms
of $\eta$ along the boundary of the worldsheet. Hence it suffices to consider
boundary observables of the form:
\be
\label{boundary_observable}
{\cal O}_\alpha(\tau)=\alpha_{{\bar i}_1\dots {\bar i}_p}(\phi(\tau))\eta^{{\bar
    i}_1}(\tau)\dots \eta^{{\bar i}_p}(\tau)~~,
\ee
where $\tau$ is a point on $\partial \Sigma$. Here $\alpha:=\alpha_{{\bar
  i}_1\dots {\bar i}_p}dz^{{\bar i}_1}\wedge \dots \wedge dz^{{\bar i}_p}$ is a $(0,p)$ form
valued in $End(E)$.

Consider a collection of $m$ topological D-branes described by superbundles
$E_a$ endowed with superconnections ${\cal B}_a$, such that the
target space equations of motion are satisfied. The index $a$ runs from
$1$ to $m$. Let $\Sigma$ be a Riemann
surface with $m$ circle boundary components $C_a$, which we
endow with the orientation induced from $\Sigma$.
Choosing forms
$\alpha^{(a)}_j\in \Omega^{(0,p^{(a)}_j)}(End(E_a))$ and points
$\tau^{(a)}_1\dots \tau^{(a)}_{k_a}$ arranged in increasing cyclic order 
along $C_a$, we are interested in the correlator:
\bea
\label{boundary_corr}
&&\langle \prod_{a=1}^{m}\prod_{j_a=k_a}^{1}
{\cal O}_{\alpha_{j_a}^{(a)}}(\tau^{(a)}_{j_a}) \rangle_\Sigma:=
\int{{\cal D}[\phi]{\cal  D}[F]{\cal D}[\rho]{\cal D}[\eta]{\cal D}[\theta]}
e^{-{\tilde  S}_{bulk}}\\
&&\prod_{a=1}^{m}{\Str\left[
{\cal O}_{\alpha^{(a)}_{k_a}}(\tau_{k_a}^{(a)})
U_a(\tau^{(a)}_{k_a},\tau^{(a)}_{k_a-1})
{\cal O}_{\alpha^{(a)}_{k_a-1}}(\tau^{(a)}_{k_a-1})\dots
{\cal O}_{\alpha^{(a)}_1}(\tau_1^{(a)})
U_a(\tau^{(a)}_1,\tau^{(a)}_{k_a})\right]}~~~,\nn
\eea
where we used the "parallel supertransport" operators defined in
(\ref{parr_tr}). The integration domain in  (\ref{boundary_corr}) is
  specified by the appropriate boundary conditions on the worldsheet fields,
  which will be discussed in more detail below. 

Let us first consider a single operator insertion 
${\cal O}_\alpha$ along a circle boundary component $C$. In this case, the
relevant factor in (\ref{boundary_corr}) is:
\be
\Str[H(\tau){\cal O}_\alpha(\tau)]~~.
\ee
We wish to compute the BRST variation of this quantity. From the relation:
\be
\label{deltaH}
\delta H(\tau)=
\left[\frac{1}{2}\int_{C}{\rho^i\partial_i W}\right] H(\tau)+[H(\tau), D(\tau)+A_{\bar
i}(\tau)\eta^{\bar i}(\tau)]~~
\ee
we obtain:
\be
\label{deltaStr}
\delta \Str[H(\tau){\cal O}_\alpha(\tau)]=
\left[\frac{1}{2}\int_{C}{\rho^i\partial_i W}\right]\Str[H(\tau){\cal
  O}_\alpha(\tau)]+\Str[H(\tau)\delta_b {\cal
  O}_\alpha (\tau)]~~,
\ee
where:
\be
\label{delta_b}
\delta_b{\cal O}_\alpha:=\delta {\cal O}_\alpha+[D+A_{\bar
i}\eta^{\bar i}, {\cal O}_\alpha]~~.
\ee
Using the target space equations of motion, one easily checks that
\footnote{Indeed, one has:
\be
\delta_b^2{\cal O}=\frac{1}{2}[F_{{\bar i}{\bar j}},{\cal O}]+\eta^{\bar
  i}[\nabla_{\bar i}D,{\cal O}]+[D^2,{\cal O}]~~.
\ee}
$\delta_b$ squares to zero, so that it plays the role of an `effective' BRST operator
in the boundary sector. Notice that $\delta_b$ arises naturally due to the
second term in the BRST variation (\ref{deltaH}) of $H(\tau)$. 
Using (\ref{delta_b}), we find the relation:
\be
\label{delta_b_loc}
\delta_b{\cal O}_\alpha={\cal O}_{{\bar D}\alpha}
\ee
where ${\bar D}={\bar D}_{\cal B}$ is the Dolbeault operator on $\Omega^{(0,*)}(End(E))$
twisted by the superconnection ${\cal B}$.

It is not hard to generalize (\ref{deltaStr}) to the case of $k$ 
insertions along $C$:
\bea
\label{delta_Str_2}
&&\delta \Str[{\cal O}_{\alpha_k}(\tau_k) U(\tau_k, \tau_{k-1})
\dots {\cal O}_{\alpha_1}(\tau_1)U(\tau_1,\tau_k)]=\\
&&~~~=\left[\frac{1}{2}\int_{C}{\rho^i\partial_i W}\right]
\Str[{\cal O}_{\alpha_k}(\tau_k) U(\tau_k, \tau_{k-1})
\dots {\cal O}_{\alpha_1}(\tau_1)U(\tau_1,\tau_k)]+\nn\\
&& ~~~~~~~+\sum_{j=1}^{k}{\Str[
{\cal O}_{\alpha_k}(\tau_k)\dots 
U(\tau_{j+1},\tau_j)\delta_b{\cal O}_{\alpha_j}(\tau_j)U(\tau_j,\tau_{j-1})
\dots {\cal O}_{\alpha_1}(\tau_1)U(\tau_1,\tau_k) ]}~~.\nn
\eea
Applying this to (\ref{boundary_corr}), we find that the BRST variation of
$e^{-{\tilde S}_{bulk}}$ is canceled by the first contribution in
(\ref{delta_Str_2}), summed over circle boundary components. This gives:
\bea
\label{deltaStr_final}
&& \delta \left(e^{-{\tilde  S}_{bulk}}\prod_{a=1}^{m}{\Str\left[
{\cal O}_{\alpha^{(a)}_{k_a}}(\tau_{k_a}^{(a)})
U(\tau^{(a)}_{k_a},\tau^{(a)}_{k_a-1})
\dots {\cal O}_{\alpha^{(a)}_1}(\tau_1^{(a)})
U(\tau^{(a)}_1,\tau^{(a)}_{k_a})\right]}\right)=~~~\\
&=&e^{-{\tilde S}_{bulk}}\sum_{a=1}^m\sum_{j_a=1}^{k_a}
\Str\left[{\cal O}_{\alpha^{(a)}_{k_a}}(\tau_{k_a}^{(a)})
U(\tau^{(a)}_{k_a},\tau^{(a)}_{k_a-1})
\dots \delta_b {\cal O}_{\alpha^{(a)}_{j_a}}(\tau_{j_a}^{(a)})
\dots {\cal O}_{\alpha^{(a)}_1}(\tau_1^{(a)})
U(\tau^{(a)}_1,\tau^{(a)}_{k_a})\right]~~.\nn
\eea
Equation (\ref{deltaStr_final}) replaces the more familiar formula known from the open
topological sigma model. Unlike the sigma model case, the left hand side 
includes the factor $e^{-{\tilde S}_{bulk}}$, because its BRST variation does
not vanish separately. Equation (\ref{deltaStr_final}) implies that the 
correlator of $\delta_b$-closed boundary observables only depends on their
$\delta_b$-cohomology class, and in particular such a correlator vanishes if
one of the boundary observables is $\delta_b$-exact. Remember that ${\tilde S}_{bulk}={\tilde
  S}_B+S_W$.
BRST closure of ${\tilde S}_B$ implies that (\ref{deltaStr_final}) is equivalent with:
\bea
\label{deltaStr_modified}
&& \delta \left(e^{-S_W}\prod_{a=1}^{m}{\Str\left[
{\cal O}_{\alpha^{(a)}_{k_a}}(\tau_{k_a}^{(a)})
U(\tau^{(a)}_{k_a},\tau^{(a)}_{k_a-1})
\dots {\cal O}_{\alpha^{(a)}_1}(\tau_1^{(a)})
U(\tau^{(a)}_1,\tau^{(a)}_{k_a})\right]}\right)=~~~\\
&=&e^{-S_W}\sum_{a=1}^m\sum_{j_a=1}^{k_a}
\Str\left[{\cal O}_{\alpha^{(a)}_{k_a}}(\tau_{k_a}^{(a)})
U(\tau^{(a)}_{k_a},\tau^{(a)}_{k_a-1})
\dots \delta_b {\cal O}_{\alpha^{(a)}_{j_a}}(\tau_{j_a}^{(a)})
\dots {\cal O}_{\alpha^{(a)}_1}(\tau_1^{(a)})
U(\tau^{(a)}_1,\tau^{(a)}_{k_a})\right]~,\nn
\eea
a fact which will be used in Section \ref{loc_boundary}.

\paragraph{Observation} It is easy to extend the discussion above by including
boundary condition changing observables, which in the present context have the
form (\ref{boundary_observable}), but with $\alpha$ an element of 
$\Omega^{(0,p)}(Hom(E_a,E_b))$. In this case, the operator 
(\ref{delta_b}) is replaced by:
\be
\delta_b{\cal O}_\alpha:=\delta {\cal O}_\alpha+
(D^{(b)}+A^{(b)}_{\bar i}\eta^{\bar i}){\cal O}_\alpha
-(-1)^{\rk\alpha}{\cal O}_\alpha (D^{(a)}+A^{(a)}_{\bar i}\eta^{\bar i})
\ee
and ${\bar D}$ in relation (\ref{delta_b_loc}) becomes the Dolbeault operator
on $\Omega^{(0,*)}(Hom(E_a,E_b))$, twisted by the superconnections 
${\cal B}_a$ and ${\cal B}_b$.

\section{Localization formula for boundary correlators on the disk}
\label{loc_boundary}

We next discuss localization in the boundary sector. As for the bulk, we will
proceed by localizing on sigma model zero-modes, thereby extracting a
two-parameter family of localization formulae. The first index of this 
family is the bulk parameter $\lambda$ of Section 3, while second parameter
$\mu$ is associated with the last term in (\ref{M_dec}). These two parameters 
measure the area and circumference length of a worldsheet with disk topology.  Each pair
$(\lambda,\mu)$ defines a localization picture, and a certain
off-shell representation of the boundary trace of \cite{Moore_Segal, Moore,
  CIL1}. As we shall see below, 
the various pictures are again related by a homotopy flow, and in particular
the various representations of the boundary trace agree when reduced to the cohomology of
$\delta_b$. In this approach, the appropriate 
generalization of the residue representation of \cite{Kap2}
is recovered in the limit $\lambda\rightarrow
+\infty$ with $\mu=0$. 

The boundary conditions induced by the coupling (\ref{Z}) can be
extracted by studying the Euler-Lagrange variations of the non-local action 
$S_{eff}={\tilde S}_{bulk}-\ln {\cal U}$. These conditions are given
explicitly in Appendix \ref{BCS}, where we also show that they are BRST
invariant modulo the equations of motion for the auxiliary fields $F$. 
The disk correlator of a collection of boundary observables ${\cal
  O}_{\alpha_1}(\tau_1)\dots {\cal O}_{\alpha_k}(\tau_k)$ (with $\tau_1\dots
\tau_k$ arranged in increasing cyclic order along the boundary) 
is obtained by performing the relevant path integral while imposing 
the  boundary conditions, which  cut out a subset ${\cal C}$ in field
configuration space:
\bea
\label{disk_correlator}
&&\langle {\cal O}_{\alpha_k}(\tau_k)\dots {\cal O}_{\alpha_1}(\tau_1)\rangle_{\rm disk}=\\
&&~~~=\int_{\cal C}{{\cal D}[\phi]{\cal D}[{\tilde F}]
{\cal D}[\eta]{\cal D}[\theta]{\cal D}[\rho]~e^{-{\tilde
    S}_{bulk}}\Str[{\cal O}_{\alpha_k}(\tau_k) U(\tau_k, \tau_{k-1})
\dots {\cal O}_{\alpha_1}(\tau_1)U(\tau_1,\tau_k)]}~~.\nn
\eea
In this section, we assume that ${\cal O}_{\alpha_j}$ are $\delta_b$-closed:
\be
\label{alpha_closure}
\delta_b{\cal O}_{\alpha_j}=0\Longleftrightarrow {\bar D}_{\cal B}\alpha_j=0~~.
\ee

\subsection{Localization on $B$-model zero-modes}

Remember that ${\tilde S}_{bulk}={\tilde S}_B+S_W$, where ${\tilde S}_B=\delta
V_B$ is BRST exact. As in Section \ref{loc_bulk}, this allows us to replace 
${\tilde S}_{bulk}$ by $t{\tilde S}_B+S_W$, without changing the value of the
correlator (\ref{disk_correlator}). Here $t$ is a complex variable with
positive real part. Invariance of (\ref{disk_correlator}) under changes in $t$ 
follows by differentiation with respect to this parameter upon using 
$\delta_b$-closure of ${\cal O}_{\alpha_j}$ and equation (\ref{deltaStr_modified})
\footnote{The path integral over ${\cal C}$ of the BRST exact term involved in
  this argument vanishes
because the boundary conditions determining ${\cal C}$ are preserved by the BRST transformations 
up to terms which vanish by the equations of motion for ${\tilde F}$ (see Appendix \ref{BCS}).}. 

We can now take the limit $\Re t\rightarrow +\infty$
to localize on the zero modes of ${\tilde S}_B$. This gives:
\be
\label{disk_correlator_loc}
\langle {\cal O}_{\alpha_k}\dots {\cal
  O}_{\alpha_1}\rangle_{\rm disk}=\langle {\cal O}_\alpha \rangle_{\rm
  disk}=N\int_{{\cal C}_0}{d\phi d\eta d\theta
e^{-\frac{A}{4}{\tilde L}_0}\Str[H_0{\cal O}_\alpha]}~~.
\ee
To arrive at this formula, 
we noticed that the dependence of $\tau_j$ disappears on zero-modes, we 
set $\alpha:=\alpha_k\wedge \dots \wedge \alpha_1$ and 
integrated out the auxiliary fields ${\tilde F}$. 
Also notice that $\delta_b {\cal O}_\alpha=0$ due to relations (\ref{alpha_closure}).
The symbol $H_0$ denotes the
restriction of the superholonomy factor $H$ to zero-modes:
\be
H_0=e^{-\frac{il}{2}k_0}~~,
\ee
where $l$ is the length of the disk's boundary and:
\be
\label{k0}
k_0:=\eta^{\bar i}\nabla_{\bar i}D^\dagger +[D,D^\dagger]=\delta_b D^\dagger~~.
\ee
The symbol ${\cal C}_0$ denotes the subset of the space of zero modes cut out
by the boundary conditions. Since we integrated out the auxiliary fields, this 
subset is strictly BRST invariant:
\be
\delta {\cal C}_0\subset {\cal C}_0~~.
\ee
Using this property as well as $\delta$-exactness (\ref{exact_tildeL_0}) of ${\tilde L}_0$
and $\delta_b$-exactness (\ref{k0}) of $k_0$, one checks
\footnote{The proof requires the identity $\delta \Str B=\Str\delta_b B$ for any
  quantity $B$ built out of zero modes. This holds because the supertrace of any
  supercommutator vanishes.}  that
(\ref{disk_correlator_loc}) is insensitive to rescalings of these
quantities, and hence can be replaced with:
\be
\label{disk_correlator_lambda_mu}
\langle {\cal O}_\alpha\rangle_{\rm disk}=N\int_{{\cal C}_0}{d\phi d\eta d\theta
e^{-\lambda{\tilde L}_0}\Str[e^{-\mu k_0}{\cal O}_\alpha]}~~,
\ee
where $\lambda$ and $\mu$ are complex numbers such that $\Re \lambda>0$. The
quantity (\ref{disk_correlator_lambda_mu}) is independent of the values of
these two parameters.

To make (\ref{disk_correlator_lambda_mu}) explicit, we must describe the
restriction to ${\cal C}_0$. The relevant boundary condition
takes the form (see Appendix \ref{BCS}):
\be
\label{eta_theta}
i(G_{i{\bar j}}\eta^{\bar j}+\theta_i){\cal U}=\Str[H(\tau)(\partial_i
D+F_{i{\bar j}}\eta^{\bar j})]\Longleftrightarrow \Str[H(\tau)(\theta_i+iV_i)]=0~~,
\ee
where:
\be
V_i:=\partial_iD+(F_{i{\bar j}}-iG_{i{\bar j}})\eta^{\bar j}~~.
\ee
Equation (\ref{eta_theta}) instructs us to replace $\theta_i$ by $-iV_i$ under
the supertrace in order to produce the desired restriction. 
To implement these constraints, we shall use the quantity:
\be
\label{Pi}
\Pi:=\frac{1}{n!}\epsilon^{i_1\dots i_n} (\theta_{i_1}
1_{End(E)}+iV_{i_1})\dots 
(\theta_{i_n} 1_{End(E)}+iV_{i_n})~~.
\ee
Consider an $End(E)$-valued function $f$ of $\theta_i$: 
\be
f(\theta_1\dots \theta_n)=\sum_{p=0}^n\sum_{1\leq i_1<\dots <i_p\leq
  n}{\theta_{i_1}\dots \theta_{i_p}f^{i_1\dots
    i_p}}=\sum_{p=0}^n\frac{1}{p!}\theta_{i_1}\dots \theta_{i_p}f^{i_1\dots i_p}~~,
\ee
where $f_{i_1\dots i_p}\in End(E)$ with $f^{i_{\sigma(1)}\dots
  i_{\sigma(p)}}=\epsilon(\sigma)f^{i_1\dots i_p}$ for all $\sigma\in
\Sigma_p$ and in the last equality we use implicit summation over $i_1\dots
i_p=1\dots n$. Here $\Sigma_p$ is the group of permutations on $p$ elements, while
$\epsilon(\sigma)$ is the signature of the permutation $\sigma$. Then one
checks the identity:
\be
\label{f_identity}
\int{d\theta_1\dots d\theta_n 
\Pi f(\theta_1\dots \theta_n)}=f(-iV_1\dots -iV_n)~~,
\ee
where the right hand side is defined by:
\bea
f(-iV_1\dots -iV_n)&:=&\sum_{p=0}^n\sum_{1\leq i_1<\dots <i_p\leq
  n}{\frac{1}{p!}\sum_{\sigma\in \Sigma_p}
\epsilon(\sigma)(-iV_{i_{\sigma(1)}})\dots (-iV_{i_{\sigma(p)}}) f^{i_1\dots
  i_p}}\nn\\
&=&\sum_{p=0}^n{\frac{1}{p!}
(-iV_{i_1})\dots (-iV_{i_p}) f^{i_1\dots
  i_p}}~~.
\eea
For later reference, we note the case $f=e^{q^i\theta_i}$, with $q^i$ some
Grassmann-odd quantities depending on $\phi$ and $\eta$. 
Then we have  $f_{i_1\dots i_p}=(-1)^{p(p+1)/2}q^{i_1}\dots q^{i_p}$
and find $f(-iV_1\dots -iV_n)=e^{-iq^i V_i}$, using the fact that
$q_j$ are mutually anti-commuting. This gives:
\be
\label{exp_identity}
\int{d\theta_1\dots d\theta_n e^{q^i\theta_i}\Pi }=\int{d\theta_1\dots d\theta_n 
\Pi e^{q^i\theta_i}}=e^{-iq^iV_i}~~.
\ee

Relation (\ref{f_identity}) shows that $\Pi$ is a sort of  `Poincare dual' of
${\cal C}_0$ on the supermanifold of
field configurations. Using (\ref{exp_identity}), this observation allows us to write
(\ref{disk_correlator_lambda_mu}) as an unconstrained integral over the space
of sphere zero-modes :
\be
\label{disk_correlator_loc_bulk}
\langle {\cal O}_\alpha\rangle_{\rm disk}=N\int{d\phi d\eta d\theta
e^{-\lambda{\tilde L}_0}\Str[e^{-\mu k_0} 
\Pi {\cal O}_\alpha ]}~~.
\ee
Employing equation (\ref{exp_identity}), we find:
\be
\label{disk_correlator_loc_boundary}
\langle {\cal O}_\alpha\rangle_{\rm disk}=N\int{d\phi d\eta
\Str[e^{-\mu k_0}e^{-\lambda{\tilde L}_0^b}{\cal O}_\alpha]}~~,
\ee
where:
\bea
\label{tildeL_0b}
{\tilde L}_0^b:={\tilde L}_0|_{\theta_i\rightarrow -iV_i}&=&-D_{\bar i}\partial_{\bar
  j}{\bar W}V^{\bar i}\eta^{\bar j}+G^{i{\bar j}}(\partial_iW)(\partial_{\bar
  j}{\bar W})=\\
&=&{\bar H}^i_{~\bar j}\eta^{\bar j}\partial_iD+\eta^{\bar
  i}\eta^{\bar j}{\bar H}^k_{~{\bar i}}F_{k{\bar j}}+G^{i{\bar j}}(\partial_iW)(\partial_{\bar
  j}{\bar W})~~.\nn
\eea
It is easy to check the relation:
\be
\delta_b V_i=\partial_i W~~,
\ee
which shows that the on-shell BRST variation (\ref{delta_theta_os}) of 
$\theta_i$ agrees with the $\delta_b$-variation of $-iV_i$:
\be
\label{delta_delta_b}
\delta_b(\theta_i 1_{End(E)}+iV_i)=0\Longleftrightarrow (\delta
\theta_i)1_{End(E)}=-i\delta_b V_i~~
\ee
(this in particular implies that $\delta \Pi=0$). Using this equation and
relation (\ref{exact_tildeL_0}), one finds that ${\tilde
  L}_0^b$ is $\delta_b$-exact:
\be
\label{tildeL_0_b_ex}
{\tilde L}_0^b=\delta_b {\tilde v}_0^b~~,
\ee
where:
\be
{\tilde v}_0^b={\tilde v}_0|_{\theta_i\rightarrow -iV_i}=V^{\bar i}\partial_{\bar i}{\bar W}~~.
\ee
Together with $\delta_b$-exactness of $k_0$, this can be used to give a direct
proof of independence of (\ref{disk_correlator_loc_boundary}) of $\lambda$ and $\mu$.

We end by mentioning some useful properties of $V_i$. It is
easy to compute the anticommutator:
\be
\label{V_commutator}
[V_i,V_j]_{-}=\partial_i\partial_jW-\delta_b[\partial_i\partial_j (D+A_{\bar k}\eta^{\bar k})]~~.
\ee
Moreover, it is not hard to check the identity:
\be
\label{VO}
[V_i, {\cal O}]=\partial_i(\delta_b{\cal O})-\delta_b(\partial_i{\cal O})
\ee
for any boundary observable ${\cal O}$ (as usual,  the quantity on the left hand
side is a supercommutator). In particular, a $\delta_b$-closed
boundary observable supercommutes with $V_i$ up to a $\delta_b$-exact term. 

\subsection{The space of boundary observables}
\label{boundary_geom}

After reduction to zero-modes, each boundary insertion ${\cal O}_\alpha$ can
be identified with the superbundle-valued differential form $\alpha$. This
amounts to setting
$\eta^{\bar i}\equiv dz^{\bar i}$, so the superalgebra ${\cal
  H}_b:=\Omega^{(0,*)}(End(E))$ provides an off-shell model for the space of boundary
excitations. Moreover, equation (\ref{delta_b_loc})
identifies the boundary BRST operator $\delta_b$ with the operator 
${\bar D}_{\cal B}={\bar \nabla}_A+{\cal D}$ acting on ${\cal H}_b$,  
where ${\bar \nabla}_A$ acts in the adjoint representation and 
${\cal D}=[D,\cdot]$. The
target space equations of motion imply that $\delta_b$ squares to zero. Thus 
${\cal H}_b$ is a differential superalgebra.  To be precise, 
we take ${\cal H}_b$ to consist of bundle-valued differential forms with at
most polynomial growth at infinity. 
 
The target space equations of motion imply the relations:
\be
{\bar \partial}_A^2={\cal D}^2={\bar \partial}_A\circ {\cal D}+{\cal D}\circ
{\bar \partial}_A=0~~,
\ee
which show that ${\cal H}_b$ is a bicomplex. Thus the boundary BRST cohomology is 
computed by a spectral sequence $E^b_*$ whose second term has the form:
\be
\label{Eb}
E^b_2=H_{\cal D}(H_{{\bar \partial}_A}({\cal H}_b))~~.
\ee

In the simple case $X=\C^n$, the holomorphic bundle  $E$ is the trivial
superbundle of type $(r_+,r_-)$ and the spectral 
sequence collapses to its second term. Then the BRST cohomology 
coincides with the cohomology of ${\cal D}$ taken in the space of square matrices of dimension
$r_++r_-$ whose entries are polynomial functions of $n$ complex variables.
This recovers the result of \cite{Kap1}.

\subsection{The boundary-bulk and bulk-boundary maps}
\label{bulk_bdry}

The equivalent expressions (\ref{disk_correlator_loc_bulk}) and 
(\ref{disk_correlator_loc_boundary}) allow us to extract an off-shell
version of the boundary-bulk map of \cite{CIL1}:
\be
f_\mu({\cal O}_\alpha)=\Str[e^{-\mu k_0}\Pi {\cal O}_\alpha]~~.
\ee
This maps ${\cal H}_b$ to ${\cal H}$ and obeys:
\be
\langle {\cal O}_\alpha\rangle_{\rm disk}=\langle f_\mu({\cal
  O}_{\alpha })\rangle_{\rm sphere}\Longleftrightarrow \Tr_b \alpha =\Tr
f_\mu(\alpha)~~,
\ee
where we identified $\alpha$ with ${\cal O}_\alpha$. Here 
$\Tr$ and $\Tr_b$ are the bulk and boundary traces determined by the 
localization formulae (\ref{correlator_final}) and (\ref{disk_correlator_loc_boundary}):
\bea
\Tr\omega~&=&\langle {\cal O}_\omega\rangle_{\rm sphere}~=\int{d\phi d\eta d\theta e^{-\lambda {\tilde L}_0}{\cal
    O}_\omega}\label{Tr}\\
\Tr_b\alpha &=&\langle  {\cal O}_\alpha\rangle_{\rm disk}=
\int{d\phi d\eta \Str[e^{-\mu k_0}e^{-\lambda {\tilde L}^b_0}{\cal
    O}_\alpha]}~~.\label{Tr_o}
\eea
As in \cite{CIL1}, we can also define a bulk-boundary map $e$ through the adjunction formula:
\be
\label{adjunction}
\Tr({\cal O}_\omega f_\mu({\cal O}_\alpha))=\Tr_b(e({\cal O}_\omega) {\cal O}_\alpha)~~.
\ee
From the relations above, we find:
\be
e({\cal O}_\omega):=e^{\lambda {\tilde L}_0^b}\int{d\theta_1\dots d\theta_n
  e^{-\lambda {\tilde L}_0}{\cal O}_\omega \Pi}~~.
\ee
This maps ${\cal H}$ to ${\cal H}_b$. 

Using (\ref{delta_delta_b}), we find that $f_\mu$ and $e$ are
compatible with the bulk and boundary BRST operators:
\bea
\delta\circ f_\mu &=& (-1)^n f_\mu\circ \delta_b ~~\label{delta_f}\\
\delta_b \circ e ~&=& e \circ \delta~~.\label{delta_e}
\eea
To prove the second equation, we used the identity:
\be
\delta\int{d\theta_1\dots d\theta_n f(\theta_1\dots
  \theta_n)}=\int{d\theta_1\dots d\theta_n \delta f(\theta_1\dots \theta_n)}~~,
\ee
which follows from (\ref{delta_theta_os}). 
Relations (\ref{delta_f}) and (\ref{delta_e}) show that 
$e$ and $f_\mu$ descend to well-defined maps $e_*$ and $f_*$ between the bulk and
boundary BRST cohomologies (the latter are the maps considered in \cite{CIL1}). 
Since $k_0$ is $\delta_b$-exact, one easily checks that $f_*$ is independent of $\mu$.

\subsection{A geometric model for the boundary trace}

As in Section \ref{loc_bulk}, we can use the identifications $\eta^{\bar
  i}\equiv dz^{\bar i}$ and $\theta_i\equiv \partial_i$ to represent our formulae in
  terms of standard geometric objects. We find:
\be
{\tilde L}_0^b\equiv{\bar H}\lrcorner(\partial D+F)+||\partial W||^2=\delta_b
  {\tilde v}_0^b~~
\ee
with:
\be
{\tilde v}_0^b={\rm grad} {\bar W}\lrcorner ~(\partial D+F)~~
\ee
and:
\be
k_0={\bar \nabla}_{\cal B}D^\dagger={\bar \nabla}_A D^\dagger+[D,D^\dagger]~~.
\ee
The disk localization formula (\ref{disk_correlator_loc_boundary}) becomes:
\be
\label{Tr_bdry_geom}
\Tr_b \alpha =N\int_{X}{\Omega\wedge \str[e^{-\lambda {\tilde L}_0^b}\wedge e^{-\mu
  k_0} \wedge \alpha]}~~, 
\ee
while the quantity (\ref{Pi}) takes the form:
\be
\Pi=\frac{1}{n!}(\partial_1+iV_1)\wedge \dots \wedge (\partial_n+iV_n)~~,
\ee
with:
\be
V_i=\partial_i D+(F_{i{\bar j}}-iG_{i{\bar j}})dz^{\bar j}~~.
\ee
Defining $V=dz^i \otimes V_i$, we obtain:
\be
V=\partial D+ F-iG~~,
\ee
where $G:=G_{i{\bar j}}dz^{i}\otimes dz^{\bar j}$. Notice that here and above,
  $\partial D$ is defined by $\partial D=dz^i\otimes \partial_i D$ (the order
  matters since $D$ is odd).

\subsection{Boundary localization pictures and the homotopy flow}

As for the bulk sector, one can define a two-parameter semigroup of operators
acting on ${\cal H}_b$ through:
\be
U_b(\lambda, \mu)(\alpha):=e^{-\lambda {\tilde L}_0^b}\wedge e^{-\mu
  k_0} \wedge \alpha~~. 
\ee
The pair $(\lambda, \mu)$ is taken inside the domain: 
\be
\Delta_b:=\{(\lambda,\mu)\in \C^2 | \Re \lambda>0 \}~~.
\ee
Since both ${\tilde L}_0$ and ${\tilde L}_0^b$ are BRST-exact, each
$U_b(\lambda,\mu)$ is homotopy-equivalent with the identity so
this defines a homotopy flow. We let: 
\be
\Tr_b^B(\alpha):=N\int_{X}{\Omega\wedge \str(\alpha)}
\ee
denote the boundary trace of the B-twisted sigma model (viewed as a linear
functional on the off-shell state space ${\cal H}_b$). Then equation (\ref{Tr_bdry_geom})
becomes: 
\be
\Tr_b(\alpha)=\Tr_b^B(U_b(\lambda,\mu)(\alpha))~~. 
\ee
Again one can define localization pictures indexed by $\lambda$ and
$\mu$. The boundary BRST operator satisfies: 
\be
U_b(\lambda,\mu)\circ\delta_b=\delta_b \circ U_b(\lambda, \mu)~~. 
\ee

\subsection{Residue formula for boundary correlators on the disk}

As in Section \ref{loc_bulk}, we can use equation (\ref{disk_correlator_loc_bulk})
to express boundary correlators in terms of generalized residues. For
simplicity, we shall assume that the spectral sequence of Subsection \ref{boundary_geom}
collapses to its second term. Setting $\mu=0$ in (\ref{disk_correlator_loc_bulk}) gives:
\be
\label{step1}
\langle {\cal O}_\alpha\rangle_{\rm disk}=N\int{d\phi d\eta d\theta
e^{-\lambda{\tilde L}_0}\Str[\Pi {\cal O}_\alpha]}~~.
\ee
We next take the limit $\lambda\rightarrow +\infty$ with $\lambda \in \R_+$. 
As in Section \ref{loc_bulk}, this forces the integral to localize on the
critical points of $W$, while the Gaussian approximation around these points
becomes exact. Counting the powers of $\lambda$ produced by the bosonic and
fermionic Gaussian integrals, we find that the correlator vanishes unless
$\alpha=f$ with $f$ a section of $End(E)$. In this case, $\delta_b$-closure of
${\cal O}_f$ amounts to the conditions ${\bar \nabla}_Af=0$ and $[D,f]=0$, 
and counting powers of $\lambda$ shows that the only contributions which
survive in the limit come from those pieces of the factor 
$\Pi$
which are independent of $\theta$ and $\eta$. This gives:
\bea
\langle {\cal O}_\alpha\rangle_{\rm disk}&=&0~~{\rm unless}~~\alpha=f\in
End(E)~~\\
\langle {\cal O}_f\rangle_{\rm disk}&=& 
\frac{C}{n!}\int_{X}{
\Omega \frac{\Str[(i\partial D)^{\wedge n}f(z)]}{\partial_1W\dots \partial_nW}}~~,
\eea
where $C$ is the constant introduced in Section \ref{loc_bulk}. These
expressions generalize the residue formula proposed in \cite{Kap2}. Notice,
however, that the residue formula of \cite{Kap2} arises for $\mu=0$ and in the
limit $\Re \lambda\rightarrow +\infty$. The limit proposed in \cite{Kap2} 
(namely $\Re\mu\rightarrow +\infty$ with $\lambda=0$) does not suffice to
localize the model's excitations unto the critical set of $W$.

\section{Conclusions}

We gave a detailed and general discussion of localization in the bulk and boundary sectors
of B-type topological Landau-Ginzburg models. In the bulk sector, we showed
that careful reconsideration of the localization argument of \cite{Vafa_LG}
leads to an entire family of localization formulae, parameterized by a complex
number $\lambda$ of positive real part. When real, this parameter 
measures the area of worldsheets with $S^2$ topology.
The various "localization pictures" are related by a "homotopy flow"
(a semigroup of operators homotopic to the identity), which implements
rescalings of this area. The generalized
localization argument leads to a one-parameter family of off-shell models for
the bulk trace, extending the well-know result of \cite{Vafa_LG}. 
The later is recovered for $\Re \lambda\rightarrow +\infty$, a degenerate
limit which leads to the standard residue representation. 

In the boundary sector, a similar argument gives a family of localization
formulae parameterized by complex variables $\lambda$ and $\mu$ subject to the
condition $\Re \lambda>0$. When real, these parameters describe the area of a worldsheet with
disk topology, respectively the length of its boundary.
The boundary localization pictures are once again related by 
a semigroup of homotopy equivalences, which implements rescaling of the disk's
area and of the length of its boundary. This leads to a two-parameter family
of off-shell models for the boundary trace. We also showed that the residue
formula proposed in \cite{Kap2} arises in the limit $\lambda\rightarrow
+\infty$ with $\mu=0$, and generalizes to the set-up of \cite{Labastida,
  coupling}, which does not require constraints on the target space or on the
rank of the holomorphic superbundle describing the relevant D-brane. In
particular, this proves and generalizes the proposal of
\cite{Kap2}, though the residue representation we have found arises in a limit which differs
from previous proposals. The argument required to establish this result 
is rather subtle, due to the complicated form of the boundary conditions.

\acknowledgments{The authors thank W. Lerche for collaboration in a related
  project. C.~I.~L. thanks A. Kapustin and K. Hori for stimulating
  conversations, and A. Klemm for support and interest in his work. }

\appendix

\section{Euler-Lagrange variations and boundary conditions}
\label{BCS}

Let us consider the Euler-Lagrange variations for our model.
It is not hard to compute the variations of the bulk action:
\bea
\delta_\theta {\tilde S}_{bulk}&=&\frac{i}{2}\int_{\Sigma}{d^2\sigma \sqrt{g}
\left[\varepsilon^{\alpha\beta}D_\beta\rho^i_\alpha+\frac{1}{2}G^{i{\bar
      k}}D_{\bar
  k}\partial_{\bar j} {\bar W}\eta^{\bar j}\right]\delta \theta_i}~~\label{bulk_EL_vars1}
\\
\delta_\eta {\tilde S}_{bulk}&=&\frac{1}{2}\int_{\Sigma}{d^2\sigma \sqrt{g}
\left[g^{\alpha\beta} G_{{\bar i}j}D_\alpha\rho^j_\beta-\frac{i}{2}D_{\bar
  i}\partial_{\bar j} {\bar W}\theta^{\bar j}\right]\delta\eta^{\bar
i}}-\frac{1}{2}\int_{\partial\Sigma}{d\tau G_{{\bar i}j}(\rho_n^j+i\rho_0^j)\delta\eta^{\bar
i}}~~~~~~~~~~\label{bulk_EL_vars2}\\
\delta_\rho {\tilde S}_{bulk}&=&\frac{1}{2}\int_{\Sigma}{d^2\sigma \sqrt{g}
\left[G_{i{\bar j}}
(i\varepsilon^{\alpha\beta}D_\beta\theta^{\bar
    j}+g^{\alpha\beta}D_\beta\eta^{\bar j})
-\frac{1}{2}\varepsilon^{\alpha\beta}D_i\partial_j W
  \rho_\beta^j\right]\delta \rho_\alpha^i}~~\nn\\
&+&\frac{i}{2}\int_{\partial\Sigma}{d\tau
G_{i{\bar j}}(\eta^{\bar j}+\theta^{\bar j})\delta \rho_0^i}~~\label{bulk_EL_vars3}\\
\delta_\phi {\tilde S}_{bulk}&=&
\int_{\Sigma}{d^2\sigma \sqrt{g}\left[
-G_{{\bar i}j}\Delta\phi^j-\frac{i}{4}D_{\bar i}D_{\bar
      j}\partial_{\bar k}{\bar W}\theta^{\bar j}\eta^{\bar
      k}+\frac{i}{2}D_{\bar i}\partial_{\bar j}{\bar W}{\tilde F}^{\bar j}
\right]\delta \phi^{\bar i}}\nn\\
&+&\int_{\Sigma}{d^2\sigma \sqrt{g}\left[
-G_{i{\bar j}}\Delta\phi^{\bar j}-\frac{i}{2}D_i\partial_j W {\tilde F}^j
+\frac{1}{8}\varepsilon^{\alpha\beta}D_iD_j\partial_k W \rho_\alpha^j\rho_\beta^k
\right]\delta \phi^i}\nn\\
&+&\int_{\partial\Sigma}{d\tau G_{{\bar i}j}(\partial_n\phi^j+i{\dot \phi}^j)\delta
  \phi^{\bar i}}+\int_{\partial\Sigma}{d\tau G_{i{\bar
      j}}(\partial_n\phi^{\bar j}-i{\dot \phi}^{\bar j})\delta
  \phi^i}~~.\label{bulk_EL_vars4}
\eea
For the boundary coupling ${\cal U}=Str H(0)$, we find:
\be
\delta {\cal U}=-\Str[H(0)I_C(\delta M)]~~,
\ee
where:
\be
I_C(\delta M)=\int_{0}^l{d\tau U(\tau)^{-1}\delta M(\tau) U(\tau)}~~.
\ee
For $\delta_M$ we substitute the Euler-Lagrange variations:
\bea
\delta_\theta M&=&0\\
\delta_\eta M&=&-\left(\frac{1}{2}F_{{\bar
      i}j}\rho_0^j+\frac{i}{2}\nabla_{\bar i}D^\dagger\right)\delta\eta ^{\bar i}\\
\delta_\rho M&=&-\frac{1}{2}(\partial_iD+F_{i{\bar j}}\eta^{\bar j})\delta \rho_0^i~~
\eea
and: 
\be
U^{-1}\delta_\phi M U=\frac{d}{d\tau}(U^{-1}A_{\bar i}\delta \phi^{\bar
  i}U)+U^{-1}(S_i\delta \phi^i+S_{\bar i}\delta \phi^{\bar i})U
\ee
with:
\bea
S_i&=&F_{i{\bar j}}{\dot \phi}^{\bar j}+\frac{1}{2}\partial_iF_{{\bar
    j}k}\eta^{\bar j}\rho_0^k+\frac{1}{2}\rho_0^j\partial_i\partial_j
D+\frac{i}{2}\eta^{\bar j}[F_{i{\bar j}},D^\dagger]+\frac{i}{2}[\partial_iD,D^\dagger]\\
S_{\bar i}&=&F_{{\bar i}j}{\dot \phi}^j+\frac{1}{2}\nabla_{\bar i}F_{{\bar
    j}k}\eta^{\bar j}\rho_0^k+\frac{1}{2}\rho_0^j[F_{{\bar
    i}j},D]+\frac{i}{2}\eta^{\bar j}\nabla_{\bar i}\nabla_{\bar j}D^\dagger
+\frac{i}{2}[D,\nabla_{\bar i}D^\dagger]~~.
\eea
This gives:
\bea
\delta_\theta {\cal U}&=&0\\
\delta_\eta {\cal
  U}&=&\int_{\partial\Sigma}{d\tau \Str[H(\tau)(\frac{i}{2}\nabla_{\bar
    i}D^\dagger+\frac{1}{2}F_{{\bar i}j}\rho_0^j)]\delta\eta^{\bar i}}\\
\delta_\rho {\cal U}&=& \frac{1}{2}\int_{\partial \Sigma}{d\tau
  \Str[H(\tau)(\partial_iD+F_{i{\bar j}}\eta^{\bar j})]\delta \rho_0^i}\\
\delta_\phi{\cal U}&=&-\int_{\partial \Sigma}{d\tau 
\left(\Str[H(\tau)S_i]\delta\phi^i+\Str[H(\tau)S_{\bar i}]\delta \phi^{\bar i}\right)}~~.
\eea
To extract the boundary conditions, we write:
\be
e^{-{\tilde S}_{bulk}}{\cal U}=e^{-S_{eff}}~~,
\ee
where $S_{eff}={\tilde S}_{bulk}-\ln {\cal U}$ is viewed as a (non-local) worldsheet action. 
Since we desire local equations of motion, the boundary contributions to
(\ref{bulk_EL_vars1}-\ref{bulk_EL_vars4}) must cancel the variation of $\ln {\cal U}$:
\be
{\cal U} \delta{\tilde S}_{bulk}=\delta {\cal U}~~.
\ee 
Imposing this requirement, we find the boundary conditions:
\bea
G_{{\bar i}j}(\rho_n^j+i\rho_0^j){\cal U}&=&-\Str H(\tau)(i\nabla_{\bar i}D^\dagger +
F_{{\bar i}j}\rho_0^j)\label{bc1}\\
i(G_{i{\bar j}}\eta^{\bar j}+\theta_i){\cal U}&=&
\Str H(\tau)(\partial_i D+F_{i{\bar j}}\eta^{\bar j})\label{bc2}\\
G_{{\bar i}j}(\partial_n\phi^j+i{\dot \phi}^j){\cal U} &=&
-\Str [H(\tau)S_{\bar i}]\label{bc3}\\
G_{i{\bar j}}(\partial_n\phi^{\bar j}-i{\dot \phi}^{\bar j}){\cal U}&=&-
\Str [H(\tau)S_i]~~.\label{bc4}
\eea
The Euler-Lagrange equations can be read off from the bulk contributions to 
(\ref{bulk_EL_vars1}-\ref{bulk_EL_vars4}):
\bea
\epsilon^{\alpha\beta}D_\alpha \rho^i_\beta&=&\frac{1}{2}G^{i{\bar j}}D_{\bar
  j}\partial_{\bar k}{\bar W}\eta^{\bar k}\\
g^{\alpha\beta}D_\alpha\rho_\beta^i&=&\frac{i}{2}G^{i{\bar j}}D_{\bar j}\partial_{\bar
  k}{\bar W}\theta^{\bar k}\\
i\epsilon^{\alpha\beta}D_\beta \theta_i+g^{\alpha\beta}G_{i{\bar j}}D_\beta\eta^{\bar
  j}&=&\frac{1}{2}\epsilon^{\alpha\beta}D_i\partial_jW
\rho_\beta^j~~.
\eea

It is not hard to see that the boundary conditions are BRST invariant modulo
the equations of motion for $F$. For simplicity, we explain this for 
condition (\ref{bc2}), which is of interest in Section
\ref{loc_boundary}. Starting with (\ref{bc2}),
one easily computes the BRST variations of the left and right hand sides:
\bea
\delta(LHS)&=&2iG_{i{\bar j}}{\tilde F}^{\bar j}{\cal U}+i(G_{i{\bar j}}\eta^{\bar
    j}+\theta_i)\left[\frac{1}{2}\int_{\partial\Sigma}\rho^i \partial_i W\right]{\cal
    U}~~\\
\delta(RHS)&=&(\partial_i W){\cal U}+
\left[\frac{1}{2}\int_{\partial\Sigma}\rho^k \partial_k W\right]
\Str[H(\tau)(\partial_iD+F_{i{\bar
    j}}\eta^{\bar j})]~~.
\eea
The two variations obviously agree if one uses equation (\ref{bc2}),
    provided that the equation of motion ${\tilde F}^{\bar
    i}=-\frac{i}{2}G^{{\bar i}j}\partial_j W$ holds.

\end{document}